\documentclass[11pt, onecolumn]{IEEEtran}

\newcommand{\Define}{\stackrel{\triangle}{=}}

\usepackage{algorithmic}
\usepackage{algorithm}
\usepackage{multirow}
\usepackage{subfigure}
\usepackage{cite}
\usepackage{latexsym}
\usepackage{epsfig}
\usepackage{verbatim}
\usepackage{amssymb}

\usepackage{graphicx}
\usepackage{color}
\usepackage{epsfig}
\usepackage[cmex10]{amsmath}

\def\BibTeX{{\rm B\kern-.05em{\sc i\kern-.025em b}\kern-.08em
    T\kern-.1667em\lower.7ex\hbox{E}\kern-.125emX}}

\long\def\symbolfootnote[#1]#2{\begingroup
\def\thefootnote{\fnsymbol{footnote}}\footnote[#1]{#2}\endgroup}

\begin{document}

\title{{\LARGE A Novel MCMC Based Receiver for Large-Scale Uplink \\ Multiuser 
MIMO Systems }} 
\author{Tanumay Datta, N. Ashok Kumar, A. Chockalingam, and B. Sundar Rajan \\
Department of ECE, Indian Institute of Science, Bangalore-560012 
}
\date{}

\maketitle

\vspace{-5mm}
\begin{abstract}
In this paper, we propose low complexity algorithms based on Markov chain
Monte Carlo (MCMC) technique for signal detection and channel estimation on
the uplink in large scale multiuser multiple input multiple output (MIMO) 
systems with tens to hundreds of antennas at the base station (BS) and similar 
number of uplink users. A BS receiver that employs a randomized sampling method 
(which makes a probabilistic choice between Gibbs sampling and random 
sampling in each iteration) for detection and a Gibbs sampling based method for 
channel estimation is proposed. The algorithm proposed for detection alleviates 
the stalling problem encountered at high SNRs in conventional MCMC algorithm and 
achieves near-optimal performance in large systems. A novel ingredient in the 
detection algorithm that is responsible for achieving near-optimal performance at 
low complexities is the joint use of a {\it randomized MCMC (R-MCMC) strategy} 
coupled with a {\it multiple restart strategy} with an efficient restart criterion.  
Near-optimal detection performance is demonstrated for large number of BS antennas 
and users (e.g., 64, 128, 256 BS antennas/users). The proposed MCMC based channel 
estimation algorithm refines an initial estimate of the channel obtained during 
pilot phase through iterations with R-MCMC detection during data phase. In time 
division duplex (TDD) systems where channel reciprocity holds, these channel 
estimates can be used for multiuser MIMO precoding on the downlink.
Further, we employ this receiver architecture in the frequency domain for 
receiving cyclic prefixed single carrier (CPSC) signals on frequency selective 
fading between users and the BS. The proposed receiver achieves performance that 
is near optimal and close to that with perfect channel knowledge.
\end{abstract}

{\em {\bfseries Keywords}} --
{\footnotesize {\em 
Large-scale multiuser MIMO, Markov chain Monte Carlo technique, Gibbs sampling, 
stalling problem, randomized sampling, multiple restarts, detection, channel 
estimation, cyclic prefixed single carrier system.
\vspace{5mm}
}}

\newpage
\section{Introduction}
\label{sec1}
Capacity of multiple-input multiple-output (MIMO) wireless channels is known
to increase linearly with the minimum of the number of transmit and receive
antennas \cite{fos98}-\cite{paul2}. Large-scale MIMO systems with tens to
hundreds of antennas have attracted much interest recently
\cite{marz}-\cite{scaling}. The motivation to consider such large-scale MIMO
systems is the potential to practically realize the theoretically predicted
benefits of MIMO, in terms of very high spectral efficiencies/sum rates, increased 
reliability and power efficiency through the exploitation of large spatial 
dimensions. Use of large number of antennas is getting recognized to be a good 
approach to fulfill the increased throughput requirements in future wireless 
systems. Particularly, large multiuser MIMO wireless systems where the base 
station (BS) has tens to hundreds of antennas and the users have one or more 
antennas are widely being investigated \cite{vtc},\cite{unlimited}-\cite{scaling}.
Communications on the uplink \cite{massive},\cite{allerton} as well as on the
downlink \cite{vtc},\cite{sriram},\cite{caire} in such large systems are of 
interest. Key issues in large multiuser MIMO systems on the downlink include 
low complexity precoding strategies and pilot contamination problem encountered 
in using non-orthogonal pilot sequences for channel estimation in multi-cell 
scenarios \cite{sriram}. In large multiuser MIMO systems on the uplink, users 
with one or more antennas transmit simultaneously to the BS with large number 
of antennas, and their signals are separated at the BS using the spatial 
signatures of the users. Sophisticated signal processing is required at the 
BS receiver to extract the signal of each user from the aggregate received signal 
\cite{tse}. Use of large number of BS antennas has been shown to improve the 
power efficiency of uplink transmissions in multiuser MIMO using linear 
receivers at the BS \cite{allerton}. Linear receivers including matched 
filter (MF) and minimum mean square error (MMSE) receivers are shown to be 
attractive for very large number of BS antennas \cite{massive}. Our focus in 
this paper is on achieving near-optimal receiver performance at the BS 
in large multiuser MIMO systems on the uplink at low complexities. The receiver 
functions we consider include {\em signal detection} and {\em channel estimation}. 
The approach we adopt for both detection as well as channel estimation is 
{\em Markov chain Monte Carlo (MCMC)} approach.

The uplink multiuser MIMO architecture can be viewed as a point-to-point MIMO 
system with co-located transmit antennas with adequate separation between them 
(so that there is no or negligible spatial correlation among them), and no 
cooperation among these transmit antennas \cite{tse}. Because of this, receiver 
algorithms for point-to-point MIMO systems are applicable for receiving uplink 
multiuser MIMO signals at the BS receiver. Recently, there has been encouraging 
progress in the development of low-complexity near-optimal MIMO receiver 
algorithms that can scale well for large dimensions 
\cite{jsac},\cite{stbc},\cite{vit}-\cite{hoehr}. These algorithms are based on 
techniques from local neighborhood search including tabu search 
\cite{jsac},\cite{stbc},\cite{vit}-\cite{r3ts}, probabilistic data association 
\cite{pda}, and message passing on graphical models including factor graphs 
and Markov random fields \cite{gm},\cite{gta},\cite{hoehr}. 

Another interesting class of low-complexity algorithms reported in the context 
of CDMA and MIMO detection is based on Markov chain Monte Carlo (MCMC) simulation 
techniques \cite{mcmc1}-\cite{gibbs}. MCMC techniques are computational techniques
that make use of random numbers \cite{mac}. MCMC methods have their roots in the
Metropolis algorithm, an attempt by physicists to compute complex integrals by
expressing them as expectations for some distribution and then estimating this
expectation by drawing samples from that distribution \cite{rob},\cite{hag}. In 
MCMC methods, statistical inferences are developed by simulating the underlying 
processes through Markov chains. By doing so, it becomes possible to reduce 
exponential detection complexity to linear/polynomial complexities. An issue 
with conventional MCMC based detection, however, is the {\em stalling problem}, 
due to which performance degrades at high SNRs \cite{mcmc2}. Stalling problem 
arises because transitions from some states to other states in a Markov chain 
can occur with very low probability \cite{mcmc2}. Our first contribution in 
this paper is that we propose an MCMC based detection algorithm that alleviates 
the stalling problem encountered in conventional MCMC and achieves near-optimal 
performance in large systems. A key idea that is instrumental in alleviating 
the stalling problem is a {\em randomized sampling strategy} that makes a 
probabilistic choice between Gibbs sampling and random sampling in each 
iteration. An efficient stopping criterion aids complexity reduction. This 
randomized sampling strategy, referred to as randomized MCMC (R-MCMC) strategy, 
is shown to achieve near-optimal performance in multiuser MIMO systems with 
16 to 256 BS antennas and same number of uplink users for 4-QAM \cite{rmcmc}. 
However, we find that this randomized sampling strategy alone is not adequate 
to achieve near-optimal performance at low complexities for higher-order QAM 
(e.g., 16-QAM, 64-QAM). We show that near-optimal performance is achieved 
in higher-order QAM also if a {\em multiple restart strategy} is performed in 
conjunction with R-MCMC. We refer to this strategy as `R-MCMC with restarts' 
(R-MCMC-R) strategy. Here again, an efficient restart criterion aids 
complexity reduction. The {\em joint use} of both randomized sampling as well as 
multiple restart strategies is found to be crucial in achieving near-optimal 
performance for higher-order QAM in large systems. To our knowledge, the 
closeness to optimal performance achieved by the proposed R-MCMC-R algorithm 
for tens to hundreds of BS antennas/users with higher-order QAM has not been 
reported so far using other MCMC based algorithms in the literature.

Channel estimation at the BS is an important issue in large multiuser MIMO 
systems on the uplink. While channel estimation at the BS is needed for 
uplink signal detection, in TDD systems where channel reciprocity 
holds, the estimated channel can also be used for precoding purposes on the 
downlink avoiding the need for feeding back channel estimates from the users.
Our second contribution in this paper is that we propose an MCMC based 
uplink channel estimation algorithm at the BS receiver. The algorithm employs 
Gibbs sampling to refine an initial estimate of the channel obtained during 
the pilot phase, through iterations with R-MCMC-R detection during the data 
phase. The algorithm is shown to yield good mean square error (MSE) and bit 
error rate (BER) performance in large multiuser MIMO systems (e.g., 128 BS 
antennas/users). BER Performance close to that with perfect channel knowledge 
is achieved. Finally, we employ the proposed MCMC based algorithms for 
equalization and channel estimation in frequency selective fading between the 
users and BS. Because of $i)$ their advantage of avoiding the peak-to-average
power ratio (PAPR) problem that is encountered in multicarrier systems and $ii)$ 
their good performance, cyclic prefixed single carrier (CPSC) transmissions 
\cite{sc1}-\cite{sc6} from the uplink users are considered. The proposed 
MCMC algorithms are shown to work well in receiving uplink CPSC transmissions 
in frequency selective fading as well. 

The rest of the paper is organized as follows. The uplink multiuser MIMO
system model on frequency non-selective fading is presented in Section
\ref{sec2}. The proposed R-MCMC algorithm without and with multiple 
restarts and its performance/complexity results in frequency non-selective 
fading are presented in Section \ref{sec3}. Section \ref{sec4} presents the 
proposed MCMC based channel estimation algorithm and its performance. 
Section \ref{sec5} presents the proposed receiver for frequency domain
equalization of CPSC signals and channel estimation in frequency selective 
fading. Conclusions are presented in Section \ref{sec6}.  

\section{System Model }
\label{sec2}
\begin{figure}
\center
\includegraphics[totalheight=3.5in,width=5.00in]{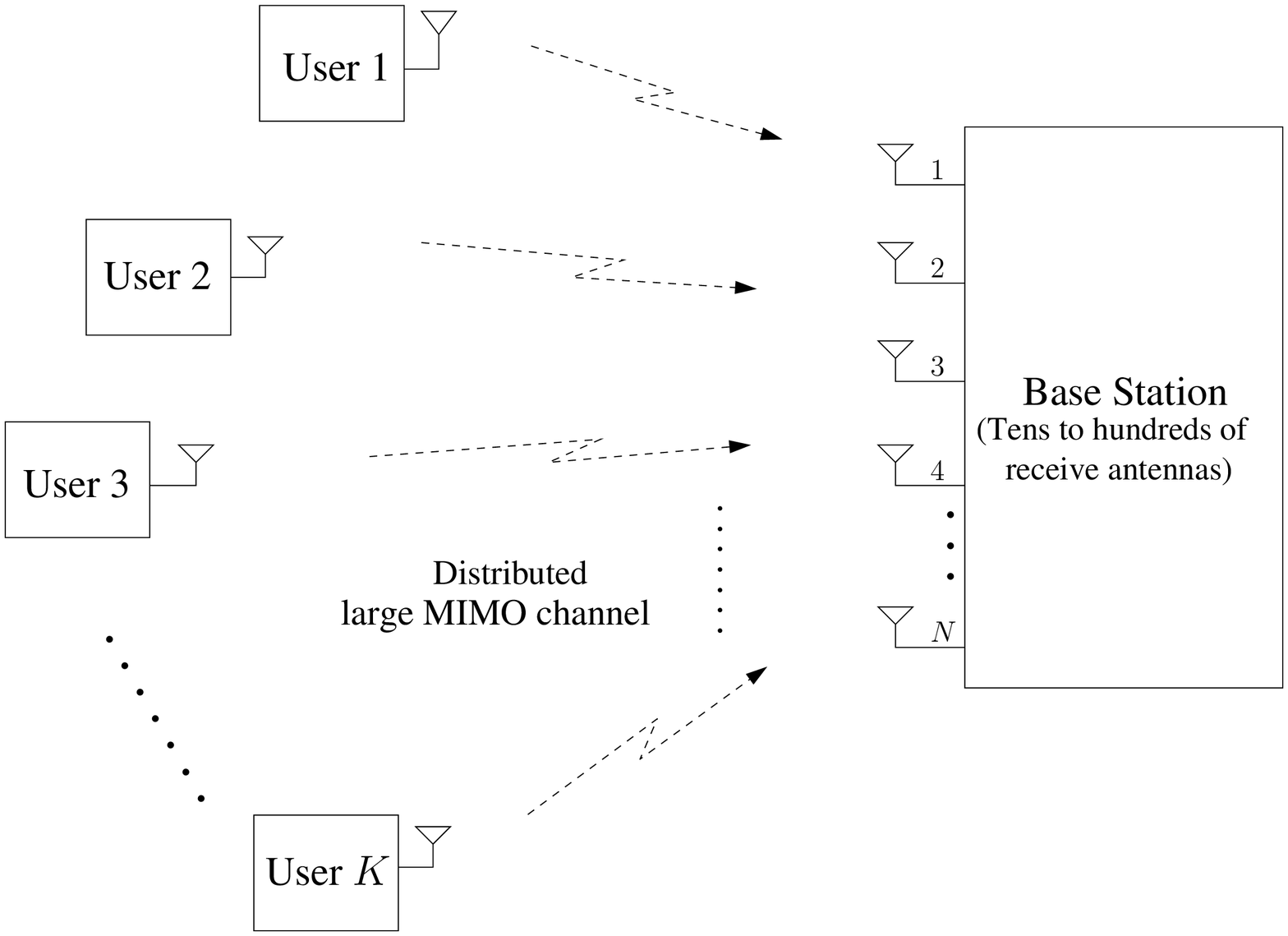}
\vspace{-2mm}
\caption{Large-scale multiuser MIMO system on the uplink.}
\label{fig1}
\end{figure}

Consider a large-scale multiuser MIMO system on the uplink consisting of a BS 
with $N$ receive antennas and $K$ homogeneous uplink users with one transmit 
antenna each, $K\leq N$ (Fig. \ref{fig1}). Extension to system model with 
non-homogeneous users where different users can have different number of 
transmit antennas is straightforward. $N$ and $K$ are in the range of tens to 
hundreds. All users transmit symbols from a modulation alphabet $\mathbb{B}$. 
It is assumed that synchronization and sampling procedures have been carried
out, and that the sampled base band signals are available at the BS receiver.
Let $x_k \in \mathbb{B}$ denote the transmitted symbol from user $k$. Let 
${\bf x}_c = [x_1, x_2, \cdots, x_K]^T$ denote the vector comprising of the 
symbols transmitted simultaneously by all users in one channel use. Let 
${\bf H}_c \in \mathbb{C}^{N\times K}$, given by
${\bf H}_c = [{\bf h}_1, {\bf h}_2, \cdots, {\bf h}_K]$,
denote the channel gain matrix, where 
${\bf h}_k = [h_{1k}, h_{2k}, \cdots, h_{Nk}]^T$ is the channel gain vector
from user $k$ to BS, and $h_{jk}$ denotes the channel gain from $k$th user 
to $j$th receive antenna at the BS. Assuming rich scattering and adequate 
spatial separation between users and BS antenna elements, $h_{jk}, \forall j$ 
are assumed to be independent Gaussian with zero mean and $\sigma_k^2$ variance
such that $\sum_{k}\sigma_k^2=K$. $\sigma_k^2$ models the imbalance in 
received powers from different users, and $\sigma_k^2=1$ corresponds to the 
perfect power control scenario. This channel gain model amounts to assuming
that the multipath fading between a user and BS is frequency non-selective.
Frequency selective fading is considered in Section \ref{sec5}. Now, the 
received signal vector at the BS in a channel use, denoted by 
${\bf y}_c \in \mathbb{C}^{N}$, can be written as
\begin{eqnarray}
{\bf y}_c & = & {\bf H}_c{\bf x}_c + {\bf n}_c,
\label{eqn1}
\end{eqnarray}
where ${\bf n}_c$ is the noise vector whose entries are are modeled as 
i.i.d. ${\mathbb C}{\mathcal N}(0,\sigma^2)$. 
We will work with the real-valued system model corresponding to (\ref{eqn1}),
given by
\begin{eqnarray}
{\bf y}_r & = & {\bf H}_r \, {\bf x}_r + {\bf n}_r, 
\label{eq_real}
\end{eqnarray}
where
${\bf x}_r \in \mathbb{R}^{2K}$, ${\bf H}_r \in \mathbb{R}^{2N\times 2K}$, 
${\bf y}_r \in \mathbb{R}^{2N}$, ${\bf n}_r \in \mathbb{R}^{2N}$ 
given by
\begin{equation}
\label{SystemModelRealDef} 
\nonumber
{\bf H}_r=\left[\begin{array}{cc}\Re({\bf H}_c) \hspace{2mm} -\Im({\bf H}_c) \\
\Im({\bf H}_c)  \hspace{5mm} \Re({\bf H}_c) \end{array}\right],
\quad
{\bf y}_r=\left[\begin{array}{c} \Re({\bf y}_c) \\ \Im({\bf y}_c) \end{array}\right],
\quad
{\bf x}_r=\left[\begin{array}{c} \Re({\bf x}_c) \\ \Im({\bf x}_c) \end{array}\right],
\quad
{\bf n}_r=\left[\begin{array}{c} \Re({\bf n}_c) \\ \Im({\bf n}_c) \end{array}\right].
\end{equation}
Dropping the subscript $r$ in (\ref{eq_real}) for notational simplicity, the 
real-valued system model is written as 
\begin{eqnarray} 
{\bf y} & = & {\bf H} {\bf x} + {\bf n}. 
\label{eqn5}
\end{eqnarray}
For a QAM alphabet ${\mathbb B}$, the elements of ${\bf x}$ will take values from 
the underlying PAM alphabet ${\mathbb A}$, i.e., ${\bf x} \in \mathbb{A}^{2K}$. 
The symbols from all the users are 
jointly detected at the BS. The maximum likelihood (ML) decision rule is  
given by
\begin{eqnarray} 
{\bf x}_{ML} & \hspace{-1mm} = & \hspace{-1mm} {\arg\min_{{\bf x} \in {\mathbb A}^{2K}}} \| {\bf y} - {\bf H}{\bf x}\|^2 
\,\, = \,\, {\arg\min_{{\bf x} \in {\mathbb A}^{2K}}} \, f({\bf x}),
\label{MLdetection}
\end{eqnarray}
where 
$f({\bf x})\Define {\bf x}^T {\bf H}^T{\bf H}{\bf x}-2{\bf y}^T{\bf H}{\bf x}$
is the ML cost. While the ML detector in (\ref{MLdetection}) is exponentially 
complex in $K$ (which is prohibitive for large $K$), the MCMC based algorithms 
we propose in the next section have quadratic complexity in $K$ and they achieve 
near-ML performance as well. 

\section{Proposed Randomized-MCMC Algorithm for Detection}
\label{sec3}
The ML detection problem in (\ref{MLdetection}) can be solved by using MCMC 
simulations \cite{mac}. We consider Gibbs sampler, which is an MCMC method 
used for sampling from distributions of multiple dimensions. In the context 
of MIMO detection, the joint probability distribution of interest is given by
\begin{eqnarray}
p(x_1,\cdots,x_{2K}|\bf{y},\bf{H})& \propto & \exp\Big(-\frac{\parallel \bf{y}-\bf{H}\bf{x}\parallel^{2}}{\sigma^{2}}\Big).  
\end{eqnarray}
We assume frequency non-selective fading and perfect knowledge of channel gain 
matrix ${\bf H}$ at the BS receiver in this section. We will relax the perfect
channel knowledge assumption by proposing a MCMC based channel estimation 
algorithm in section \ref{sec4}.  

\subsection{Conventional MCMC algorithm}
\label{sec3a}
In conventional Gibbs sampling based detection, referred to as conventional
MCMC algorithm, the algorithm starts with an 
initial symbol vector, denoted by ${\bf x}^{(t=0)}$. In each iteration of the 
algorithm, an updated symbol vector is obtained by sampling from 
distributions as follows: 
\begin{eqnarray}
{x}^{(t+1)}_1 & \thicksim & p(x_1|x^{(t)}_2,{x}^{(t)}_3,\cdots,x^{(t)}_{2K}), \nonumber \\
x^{(t+1)}_2 & \thicksim & p(x_2|x^{(t+1)}_1,x^{(t)}_3,\cdots,x^{(t)}_{2K}), \nonumber \\
x^{(t+1)}_3 & \thicksim & p(x_3|x^{(t+1)}_1,x^{(t+1)}_2,x^{(t)}_4,\cdots,x^{(t)}_{2K}), \nonumber \\
 &\vdots & \nonumber  \\
x^{(t+1)}_{2K} & \thicksim & p(x_{2K}|x^{(t+1)}_1,x^{(t+1)}_2,\cdots,x^{(t+1)}_{2K-1}).
\label{update}
\end{eqnarray}
The detected symbol vector in a given iteration is chosen to be that symbol 
vector which has the least ML cost in all the iterations up to that 
iteration.
Another MCMC algorithm that uses a temperature parameter $\alpha$ and the 
following joint distribution is presented in \cite{gibbs}: 
\begin{eqnarray}
p(x_1,\cdots,x_{2K}|\bf{y},\bf{H})& \propto & \exp\Big(-\frac{\parallel \bf{y}-\bf{H}\bf{x}\parallel^{2}}{\alpha^{2}\sigma^2}\Big).
\label{alpha}
\end{eqnarray}
The algorithm uses a fixed value of $\alpha$ in all the iterations, with the 
property that after the Markov chain is mixed, the probability of encountering 
the optimal solution is only polynomially small (not exponentially small). This 
algorithm and the conventional MCMC algorithm (which is a special case of 
$\alpha=1$) face stalling problem at high SNRs; a problem in which the BER 
performance gets worse at high SNRs \cite{mcmc2}. 

\subsection{Proposed R-MCMC algorithm}
\label{sec3b}
It is noted that the stalling problem occurs due to MCMC iterations getting
trapped in poor local solutions, beyond which the ML cost does not improve with 
increasing iterations for a long time. Motivated by this observation, we propose
a simple, yet effective, randomization strategy to avoid such traps. The key idea 
behind the proposed randomized MCMC (R-MCMC) approach is that, in each iteration,
instead of updating $x_i^{(t)}$'s as per the update rule in (\ref{update}) with 
probability 1 as in
conventional MCMC, we update them as per (\ref{update}) with probability $(1-q_i)$ 
and use a different update rule with probability $q_i=\frac{1}{2K}$. The different
update rule is as follows. Generate $|{\mathbb A}|$ probability values from uniform 
distribution as 
\begin{eqnarray*}
p(x_i^{(t)}=j) \thicksim U[0,1], \quad \forall j\in {\mathbb A} 
\end{eqnarray*}
such that
$\sum\limits_{j=1}^{|{\mathbb A}|} p(x_i^{(t)}=j) = 1$, and
sample $x_i^{(t)}$ from this generated pmf. 

\subsubsection{Proposed stopping criterion}
\label{sec3b1}
A suitable termination criterion is needed to stop the algorithm. A simple 
strategy is to terminate the algorithm after a fixed number of iterations.
But a fixed value of number of iterations may not be appropriate for all
scenarios. Fixing a large value for the number of iterations can yield good
performance, but the complexity increases with the number of iterations. To 
address this issue, we develop a dynamic stopping criterion that yields good 
performance without unduly increasing the complexity. The criterion works as 
follows. A stalling is said to have occurred if the ML cost remains unchanged 
in two consecutive iterations. Once such a stalling is identified, the algorithm 
generates a positive integer $\Theta_s$ (referred to as the {\em stalling limit}), 
and the iterations are allowed to continue in stalling mode (i.e., without 
ML cost change) up to a maximum of $\Theta_s$ iterations from the occurrence of 
stalling. If a lower ML cost is encountered before $\Theta_s$ iterations, the 
algorithm proceeds with the newly found lower ML cost; else, the algorithm
terminates. If termination does not happen through stalling limit as above, 
the algorithm terminates on completing a maximum number of iterations, MAX-ITER. 

The algorithm chooses the value of $\Theta_s$ depending on the quality of the 
stalled ML cost, as follows. A large value for $\Theta_s$ is preferred if the 
quality of the stalled ML cost is poor, because of the available potential for 
improvement from a poor stalled solution. On the other hand, if the stalled ML 
cost quality is already good, then a small value of $\Theta_s$ is preferred. 
The quality of a stalled solution is determined in terms of closeness of the 
stalled ML cost to a value obtained using the statistics (mean and variance) 
of the ML cost for the case when ${\bf x}$ is detected error-free. Note that 
when ${\bf x}$ is detected error-free, the corresponding ML cost is nothing 
but $\|{\bf n}\|^2$, which is Chi-squared distributed with $2N$ degrees of 
freedom with mean $N\sigma^{2}$ and variance $N\sigma^{4}$. We define the
quality metric to be the difference between the ML cost of the stalled 
solution and the mean of $\|{\bf n}\|^2$, scaled by the standard deviation, 
i.e., the quality metric of vector $\hat{\bf x}$ is defined as 
\begin{eqnarray}
\phi({\hat{\bf  x}})=\frac{\|{\bf y}-{\bf H}\hat{{\bf  x}}\|^2-N\sigma^{2}}{\sqrt{N}\sigma^{2}}.
\label{qual}
\end{eqnarray}
We refer to the metric in (\ref{qual}) as the {\em standardized ML cost} of 
solution vector $\hat{\bf x}$.
A large value of $\phi(\hat{\bf x})$ can be viewed as an indicator of increased 
closeness of $\hat{\bf x}$ to ML solution. Therefore, from the previous discussion,  
it is desired to choose the stalling limit $\Theta_s$ to be an increasing 
function of $\phi({\hat{\bf x}})$. For this purpose, we choose an exponential 
function of the form 
\begin{equation}
\Theta_s(\phi(\hat{{\bf x}}))=c_1\exp(\phi(\hat{{\bf x}})).
\end{equation}
Also, we allow a minimum number of iterations ($c_{min}$) following a stalling.
Based on the above discussion, we adopt the following rule to compute the 
stalling count: 
\begin{eqnarray}
\Theta_s(\hat{\bf x}) = \left\lceil \max\left(c_{min},c_1\exp\left(\phi(\hat{{\bf x}})\right)\right)\right\rceil.
\label{limit}
\end{eqnarray}
The constant $c_1$ is chosen depending upon the QAM size; a larger $c_1$
is chosen for larger QAM size. As we will see in the performance and 
complexity results, the proposed randomization in the update rule and the 
stopping criterion are quite effective in achieving low complexity as well as 
near-optimal performance. A complete listing of the proposed R-MCMC algorithm 
incorporating the randomized sampling and stopping criterion ideas is given 
in the next page.

\begin{algorithm}
\caption{Proposed randomized-MCMC algorithm}
\begin{algorithmic}[1]
\STATE {{\bf input:} ${\bf y}$, ${\bf H}$, ${\bf x}^{(0)};$\\ 
\vspace{-2mm}
${\bf{x}}^{(0)}:$ initial vector $\in {\mathbb{A}}^{2K}$; \,\,
MAX-ITER: max. \# iterations;}
\vspace{-2mm}
\STATE $t=0$; \,\, ${\bf z}={\bf x}^{(0)}$; 
\,\, $S=\{1,2,\cdots,2K\}$; \,\, $S:$ index set; \\
\vspace{-2mm}
\STATE $\beta=f({\bf{x}}^{(0)}); \quad f(.):$ ML cost function; \,\, 
$ \Theta_s(.):$ \textit{stalling limit} function;
\vspace{-2mm}
\WHILE{$t <$ MAX-ITER}
\vspace{-2mm}
\FOR{$i=1$ to $2K$}
\vspace{-2mm}
\STATE randomly choose an index $k$ from the set $S$;\\
\vspace{-2mm}
\IF{$(i \neq k)$}
\STATE {\small ${x}^{(t+1)}_i \thicksim  p(x_i| {x}^{(t+1)}_1,\cdots,{x}^{(t+1)}_{i-1},{x}^{(t)}_{i+1},\cdots,{x}^{(t)}_{2K}) $}\\
\vspace{-2mm}
\ELSE
\vspace{-2mm}
\STATE generate pmf $p(x_i^{(t+1)}=j) \thicksim U[0,1], \,\, \forall j\in {\mathbb A}$ \\
\vspace{-2mm}
\STATE {sample $x_i^{(t)}$ from this pmf} \\
\vspace{-2mm}
\ENDIF\\
\vspace{-2mm}
\ENDFOR\\
\vspace{-2mm}
\STATE {$\gamma=f({\bf{x}}^{(t+1)});$}\\
\vspace{-2mm}
\IF{$(\gamma \leq \beta)$}
\vspace{-2mm}
\STATE ${\bf z}={\bf x}^{(t+1)};$ \,\,\, $\beta=\gamma$;\\
\vspace{-2mm}
\ENDIF\\
\vspace{-2mm}
\STATE {$t=t+1;$}\\
\vspace{-2mm}
\STATE{$\beta_v^{(t)}=\beta;$}\\
\vspace{-2mm}
\IF {$\beta_v^{(t)}==\beta_v^{(t-1)}$}
\vspace{-2mm}
\STATE{calculate $\Theta_s({\bf z})$;}
\vspace{-2mm}
\IF{$\Theta_s<t$}
\vspace{-2mm}
\IF {$\beta_v^{(t)}==\beta_v^{\left( t-\Theta_s\right)}$} 
\vspace{-2mm}
\STATE goto step 29 \\
\vspace{-2mm}
\ENDIF\\
\vspace{-2mm}
\ENDIF\\
\vspace{-2mm}
\ENDIF\\
\vspace{-2mm}
\ENDWHILE\\
\vspace{-2mm}
\STATE {{\bf output:} ${\bf z}. \qquad{\bf z}:$ output solution vector}\\
\end{algorithmic}
\end{algorithm}

\subsubsection{Performance and complexity of the R-MCMC algorithm} 
\label{sec3b2}
\begin{figure}
\center
\epsfysize=9.0cm
\epsfxsize=12.00cm
\epsfbox{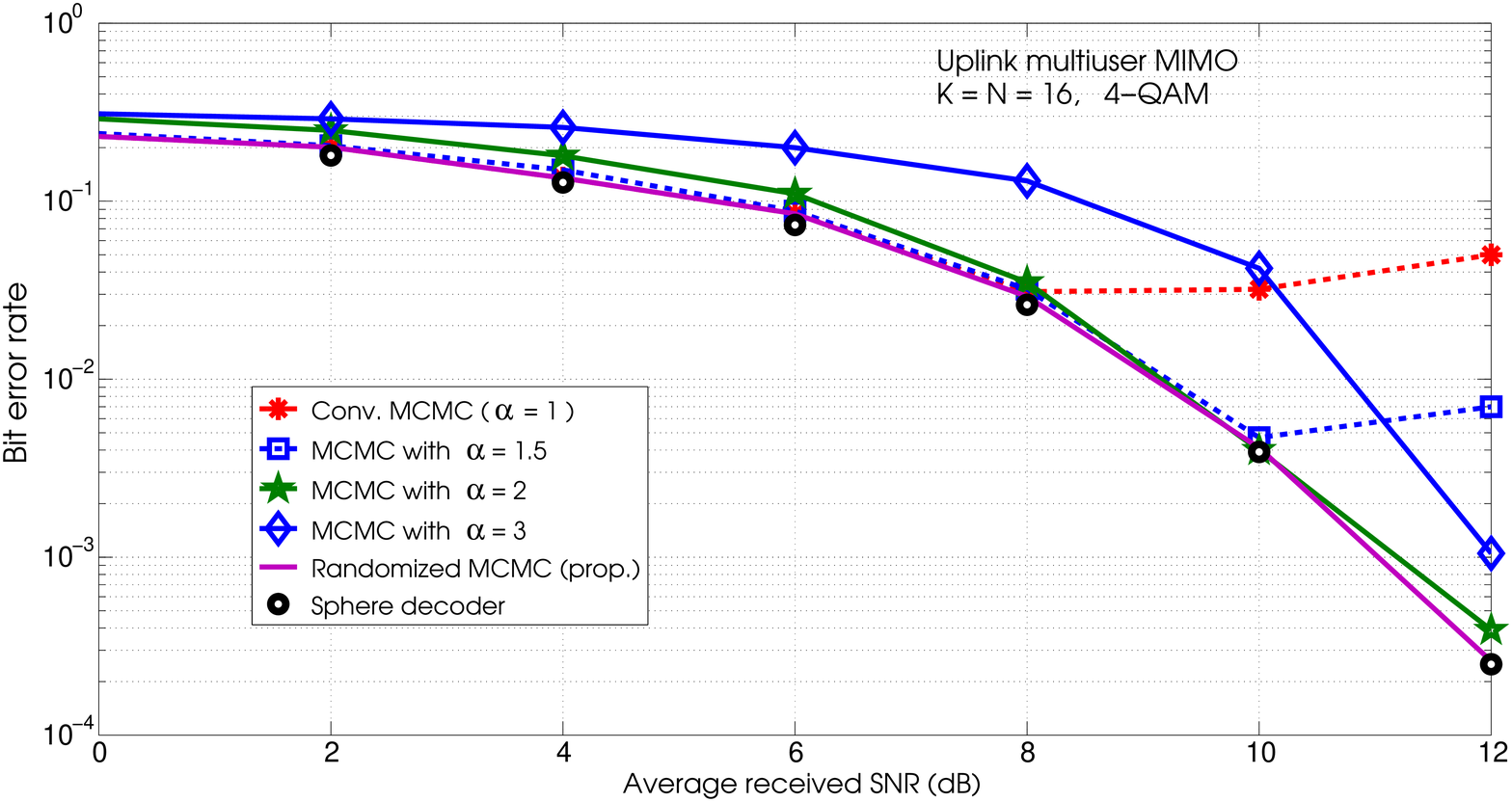}
\vspace{-6.0mm}
\caption{ 
BER performance of the proposed R-MCMC algorithm in comparison with those of 
sphere decoder and MCMC algorithm with different values of $\alpha$ in uplink
multiuser MIMO with $K=N=16$, 4-QAM, and no power imbalance. Performance of
R-MCMC is almost the same as the sphere decoder performance.  }
\vspace{-4.0mm}
\label{fig2}
\end{figure}

\begin{figure}
\center
\epsfysize=9.0cm
\epsfxsize=12.00cm
\hspace{-4mm}
\epsfbox{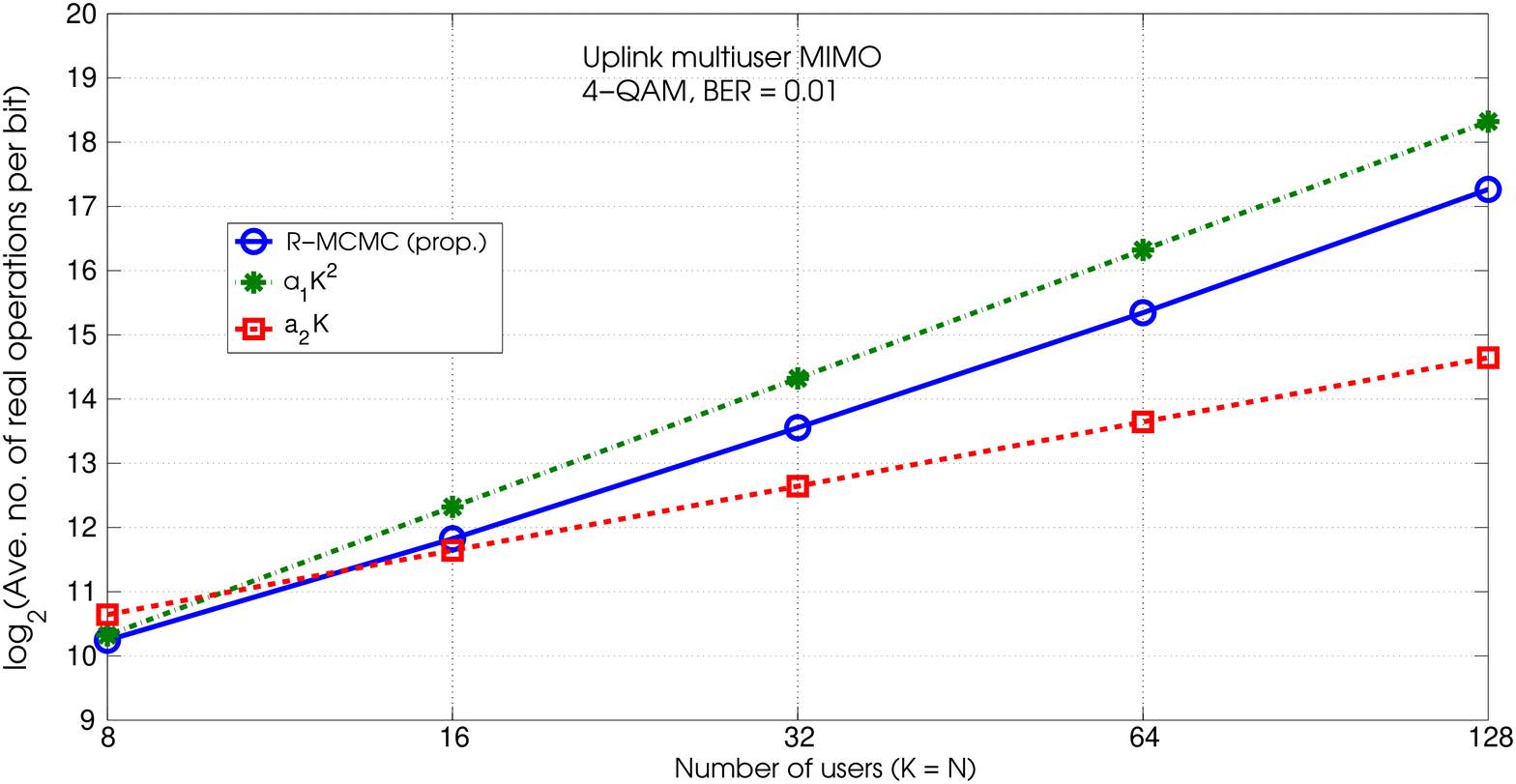}
\vspace{-6.0mm}
\caption{Complexity of the R-MCMC algorithm in average number of real 
operations per bit as a function of $K=N$ with 4-QAM and no power 
imbalance at $10^{-2}$ BER.}
\vspace{-4.0mm}
\label{fig3}
\end{figure}

\begin{figure}
\center
\epsfysize=9.0cm
\epsfxsize=12.00cm
\hspace{-4mm}
\epsfbox{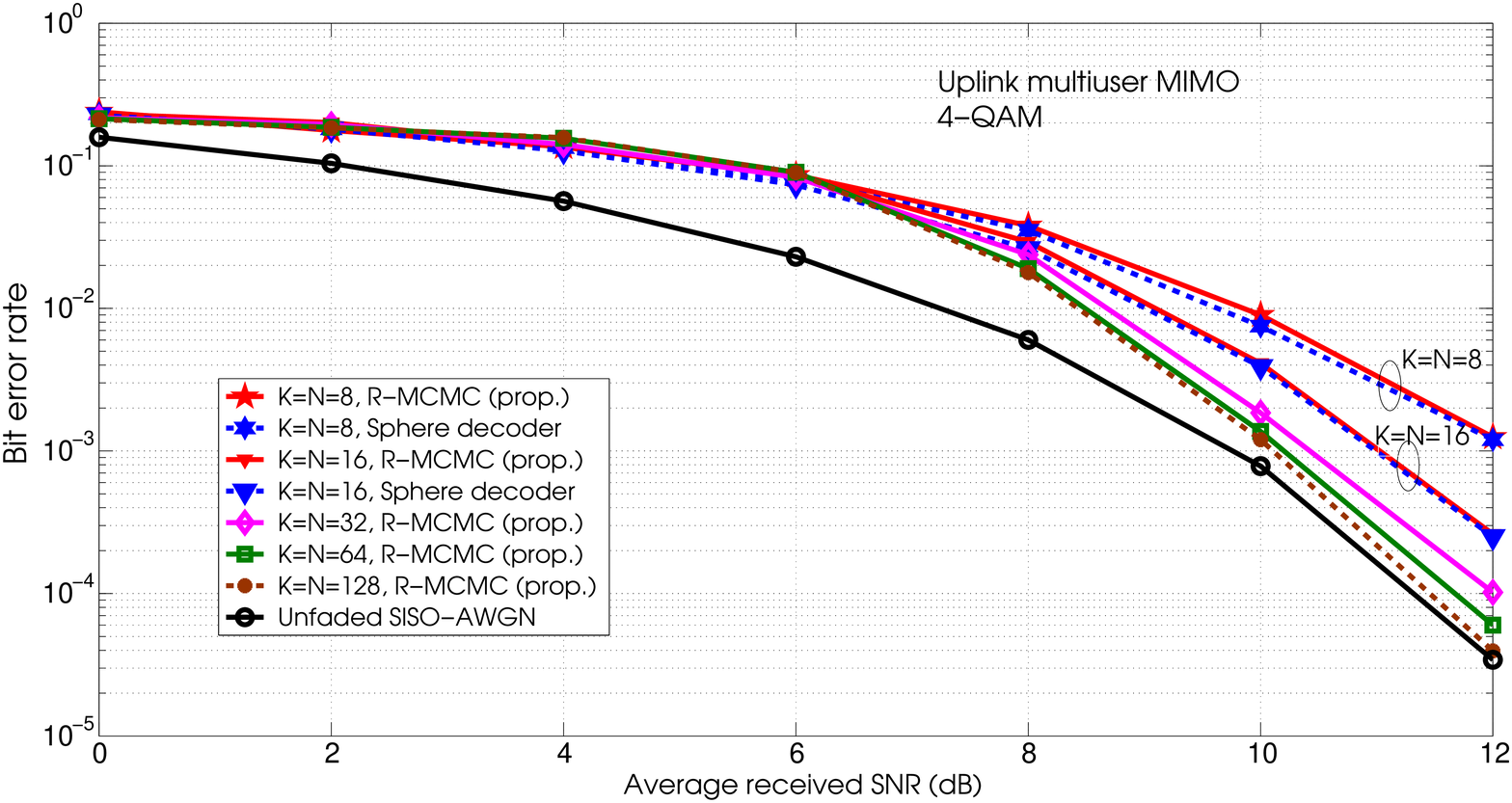}
\vspace{-6.0mm}
\caption{BER performance of the R-MCMC algorithm in uplink multiuser MIMO 
with $K=N=8, 16, 32,64, 128$, 4-QAM and no power imbalance.}
\vspace{-4.0mm}
\label{fig4}
\end{figure}

\begin{figure}
\center
\epsfysize=9.0cm
\epsfxsize=12.00cm
\hspace{-4mm}
\epsfbox{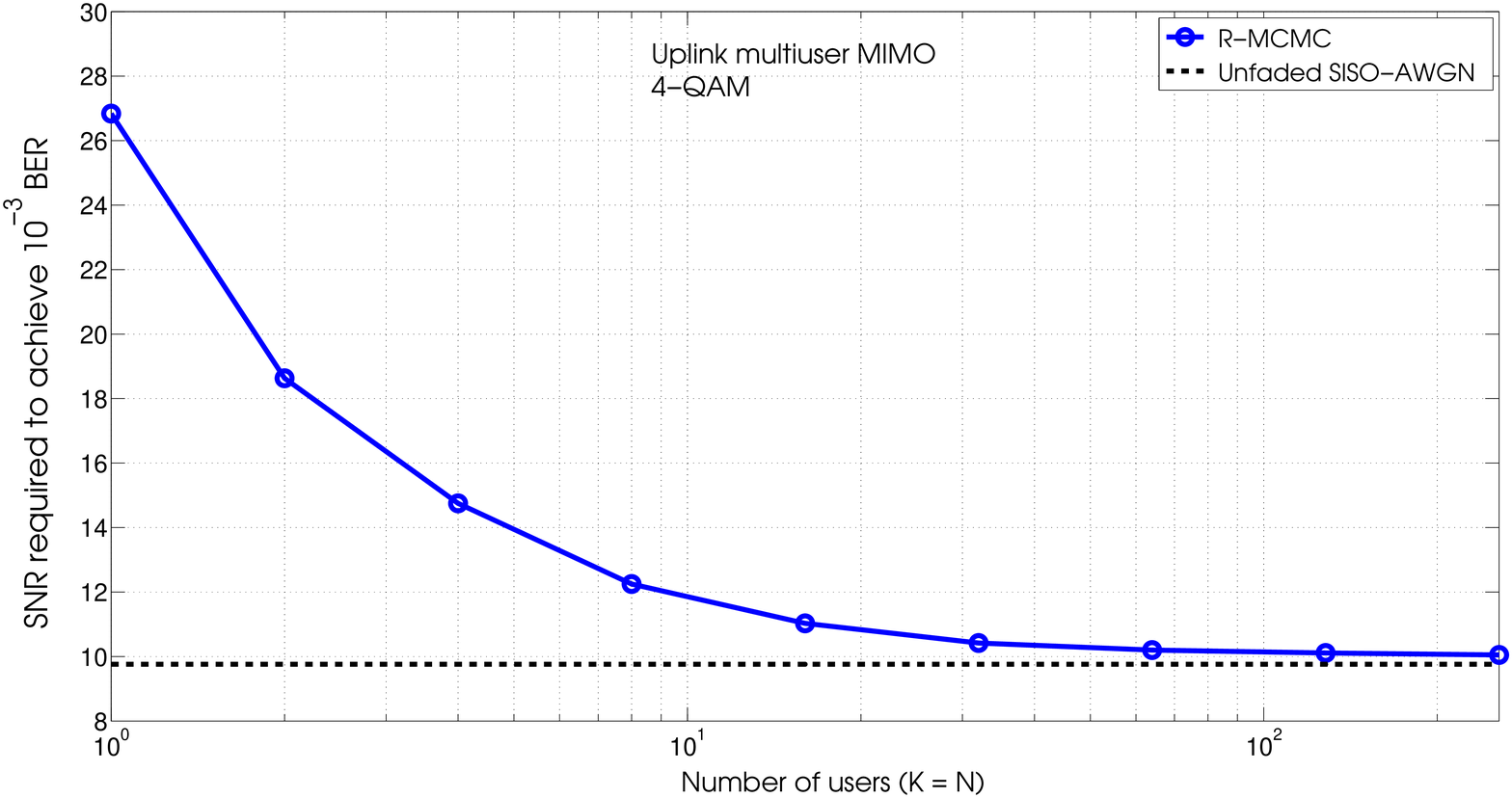}
\vspace{-6.0mm}
\caption{Average SNR required to achieve $10^{-3}$ BER as a function of number
of users $(K=N)$ in uplink multiuser MIMO with 4-QAM and no power imbalance.}
\vspace{-4.0mm}
\label{fig5}
\end{figure}

\begin{figure}
\center
\epsfysize=9.0cm
\epsfxsize=12.00cm
\hspace{-4mm}
\epsfbox{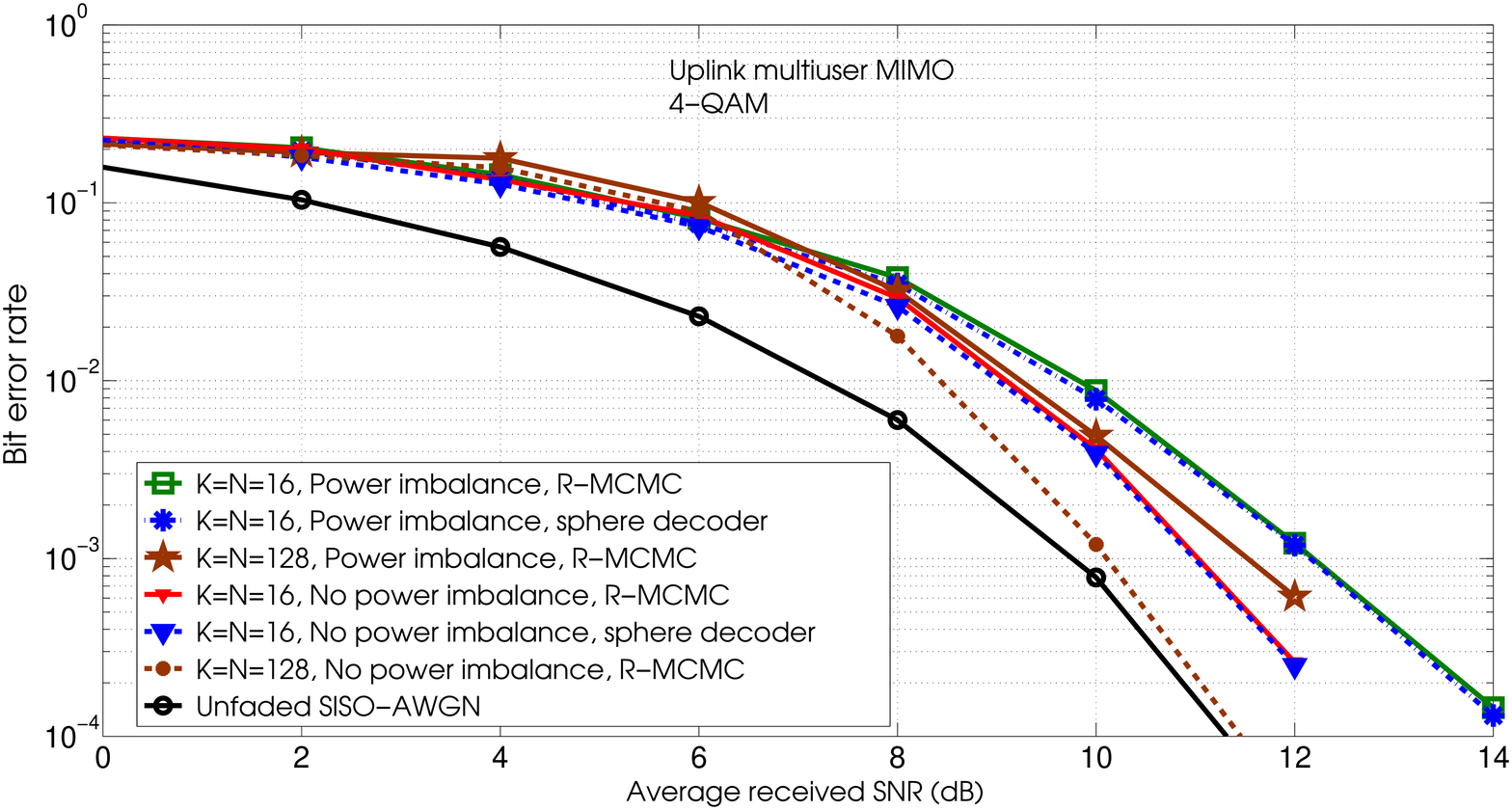}
\vspace{-6.0mm}
\caption{BER performance of the R-MCMC algorithm in uplink multiuser MIMO 
with $K=N=16,128$ and 4-QAM. Power imbalance with $\sigma_k^2$'s uniformly 
distributed between -3 dB and 3 dB.}
\vspace{-4.0mm}
\label{fig6}
\end{figure}

The simulated BER performance and complexity of the proposed R-MCMC algorithm 
in uplink multiuser MIMO systems with 4-QAM are shown in Figs. \ref{fig2} to 
\ref{fig6}. The following R-MCMC parameters are used in the simulations: 
$c_{min}=10$, $c_1=20$, MAX-ITER $= 16K$. Figures \ref{fig2} to \ref{fig5} 
are for the case where there is no imbalance in the received powers of all 
users, i.e., $\sigma_k^2=0$ dB $\forall\, k$. Perfect channel knowledge 
at the BS is assumed. The performance of R-MCMC in 
multiuser MIMO with $K=N=16$ is shown in Fig. \ref{fig2}. The performance of 
the MCMC algorithm using the distribution in (\ref{alpha}) with temperature 
parameter values $\alpha=1,1.5,2,3$ are also plotted. $16K$ iterations are used 
in the MCMC algorithm with temperature parameter. Sphere decoder performance 
is also shown for comparison. It is seen that the performance of MCMC with 
temperature parameter is very sensitive to the choice of the value of $\alpha$. 
For example, for $\alpha=1,1.5$, the BER is found to degrade at high SNRs due to 
stalling problem. For $\alpha=2$, the performance is better at high SNRs but 
worse at low SNRs. The proposed R-MCMC performs better than MCMC with temperature 
parameter (or almost the same) at all SNRs and $\alpha$ values shown. In fact, the 
performance of R-MCMC is almost the same as the sphere decoder performance. The 
R-MCMC complexity is, however, significantly lower than the sphere decoding 
complexity. While sphere decoder gets exponentially complex in $K$ at low SNRs, 
the R-MCMC complexity (in average number of real operations per bit) is only 
$O(K^2)$ as can be seen in Fig. \ref{fig3}. Because of this low complexity,
the R-MCMC algorithm scales well for large-scale systems with large values
of $K$ and $N$. This is illustrated in Fig. \ref{fig4} and \ref{fig5} where 
performance plots for systems up to $K=N=128$ and 256 are shown. While Fig. 
\ref{fig4} shows the BER as a function of SNR, Fig. \ref{fig5} shows the average 
received SNR required to achieve a target BER of $10^{-3}$ as a function of $K=N$.  
Since sphere decoder complexity is prohibitive for hundreds of dimensions, we 
have plotted unfaded single-input single-output (SISO) AWGN performance as a 
lower bound on ML performance for comparison. It can be seen that R-MCMC achieves 
performance which is very close to SISO AWGN performance for large $K=N$, e.g., 
close to within 0.5 dB at $10^{-3}$ BER for $K=N=128$ and 256. This illustrates 
the achievability of near-optimal performance using R-MCMC for large systems. 
Figure \ref{fig6} shows the BER performance in multiuser MIMO systems with 
received power imbalance among different users. The imbalance is simulated by 
choosing different $\sigma_k^2$ for different users, with $\sigma_k^2$ being 
uniformly distributed between -3 dB to 3 dB. Performance in systems with
$K=N=16$ and 128 are plotted with and without power imbalance. It is seen that 
even with power imbalance R-MCMC achieves almost the same performance as that 
of sphere decoder for $K=N=16$.

\subsection{Multi-restart R-MCMC algorithm for higher-order QAM}
\label{sec3c}
\begin{figure}
\center
\includegraphics[totalheight=9.0cm,width=12.00cm]{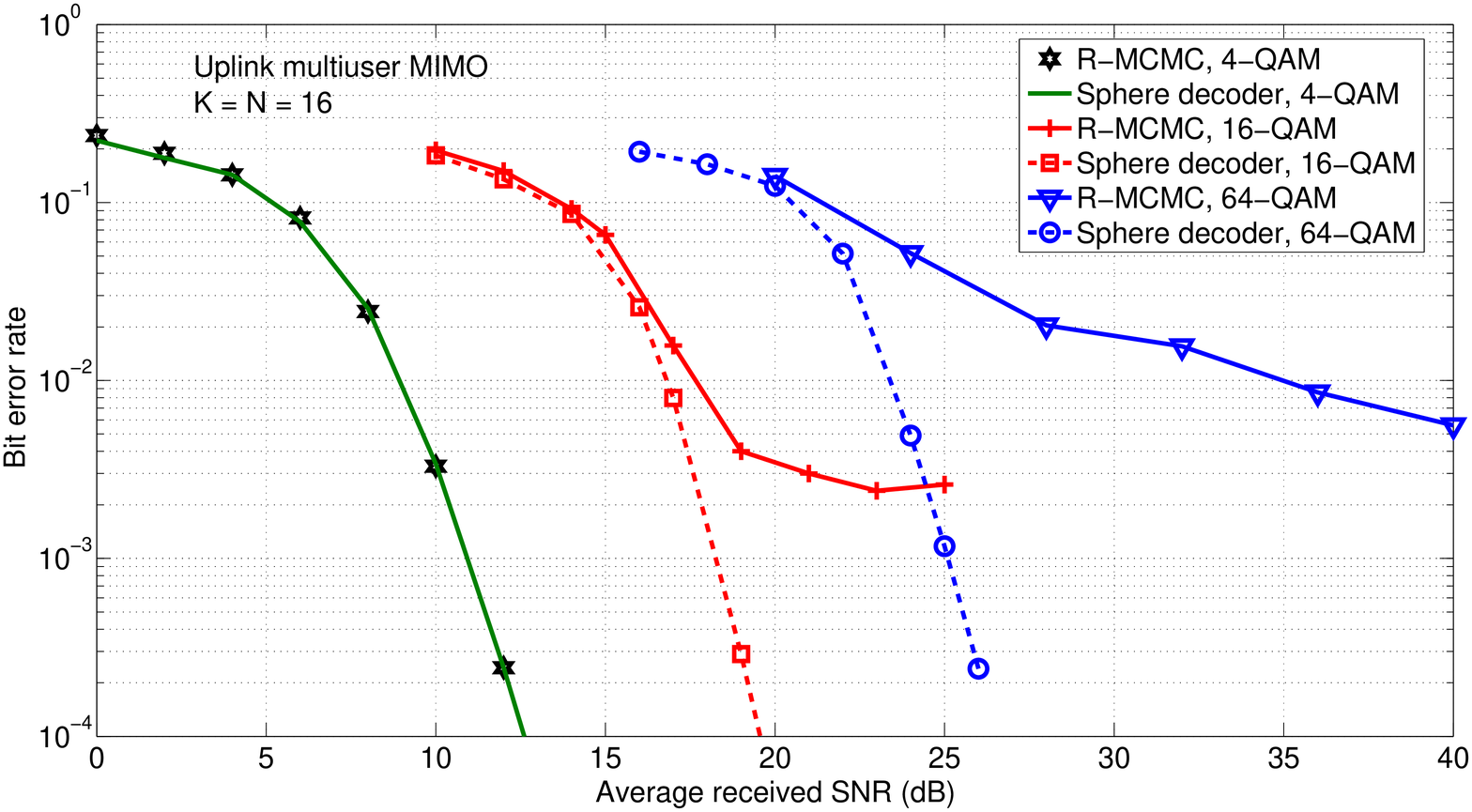}
\vspace{-6mm}
\caption{Comparison between R-MCMC performance and sphere decoder performance 
in uplink multiuser MIMO with $K=N=16$ and 4-/16-/64-QAM.} 
\vspace{-4.0mm}
\label{fig7}
\end{figure}

Although the R-MCMC algorithm is very attractive in terms of both performance 
as well as complexity for 4-QAM, its performance for higher-order QAM is far 
from optimal. This is illustrated in Fig. \ref{fig7}, where R-MCMC is seen to 
achieve sphere decoder performance for 4-QAM, whereas for 16-QAM and 64-QAM it 
performs poorly compared to sphere decoder. This observation motivates the need
for ways to improve R-MCMC performance in higher-order QAM. Interestingly, we 
found that use of multiple restarts\footnote{It is noted that multiple restarts, 
also referred to as running multiple parallel Gibbs samplers, have been tried with 
conventional and other variants of MCMC in \cite{mcmc2},\cite{mcmc4},\cite{mcmc5}. 
But the stalling problem is not fully removed and near-ML performance is not 
achieved. It turns out that restarts when coupled with R-MCMC is very 
effective in achieving near-ML performance.} coupled with R-MCMC is able 
to significantly improve performance and achieve near-ML performance 
in large systems with higher-order QAM.

\subsubsection{Effect of restarts in R-MCMC and conventional MCMC}
\label{sec3c1}
\begin{figure}
\hspace{-6mm}
\subfigure[4-QAM, SNR=11 dB]{
\includegraphics[width=3.5in,height=3.00in]{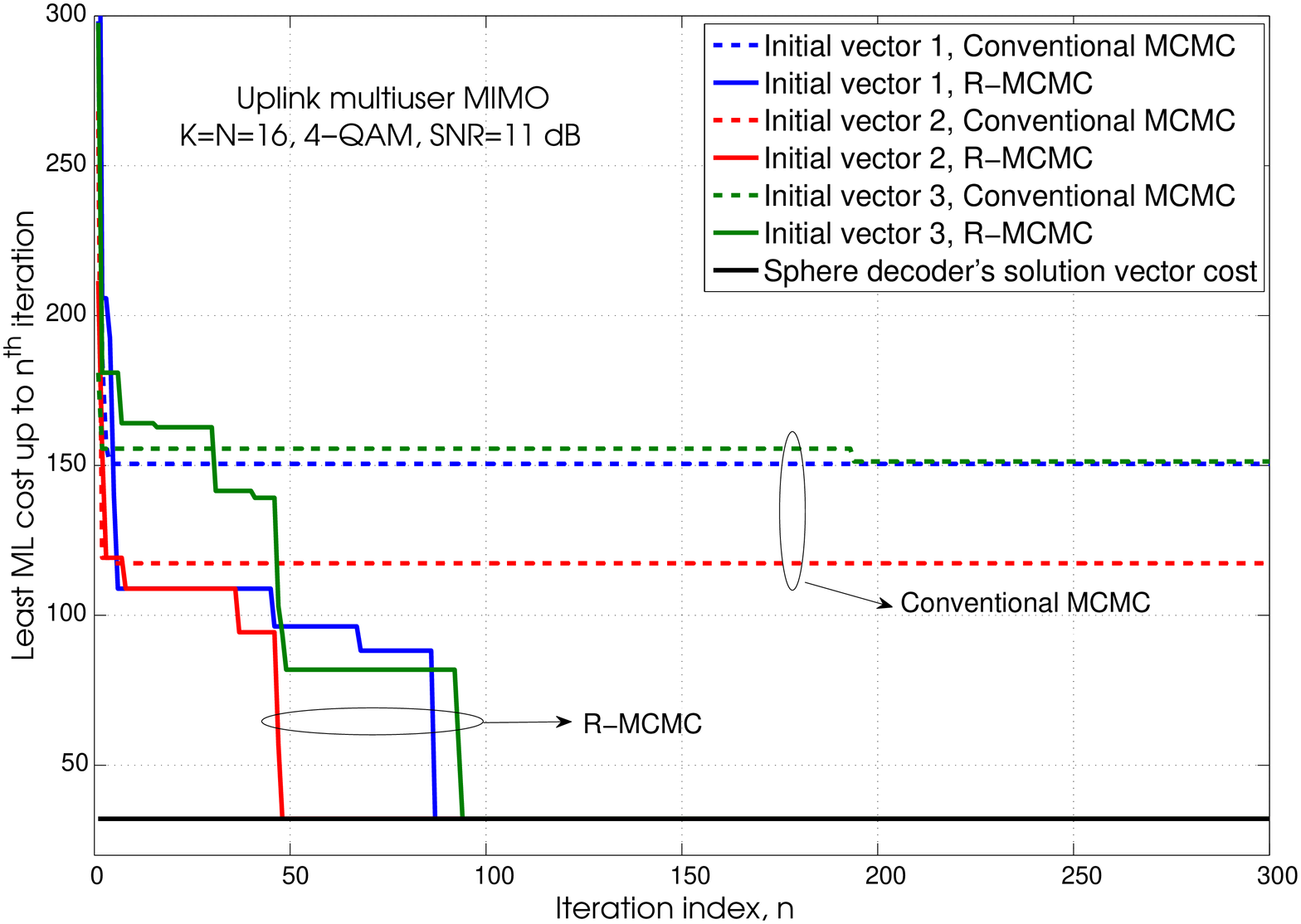}}
\hspace{-6mm}
\subfigure[16-QAM, SNR=18 dB]{
\includegraphics[width=3.5in,height=3.00in]{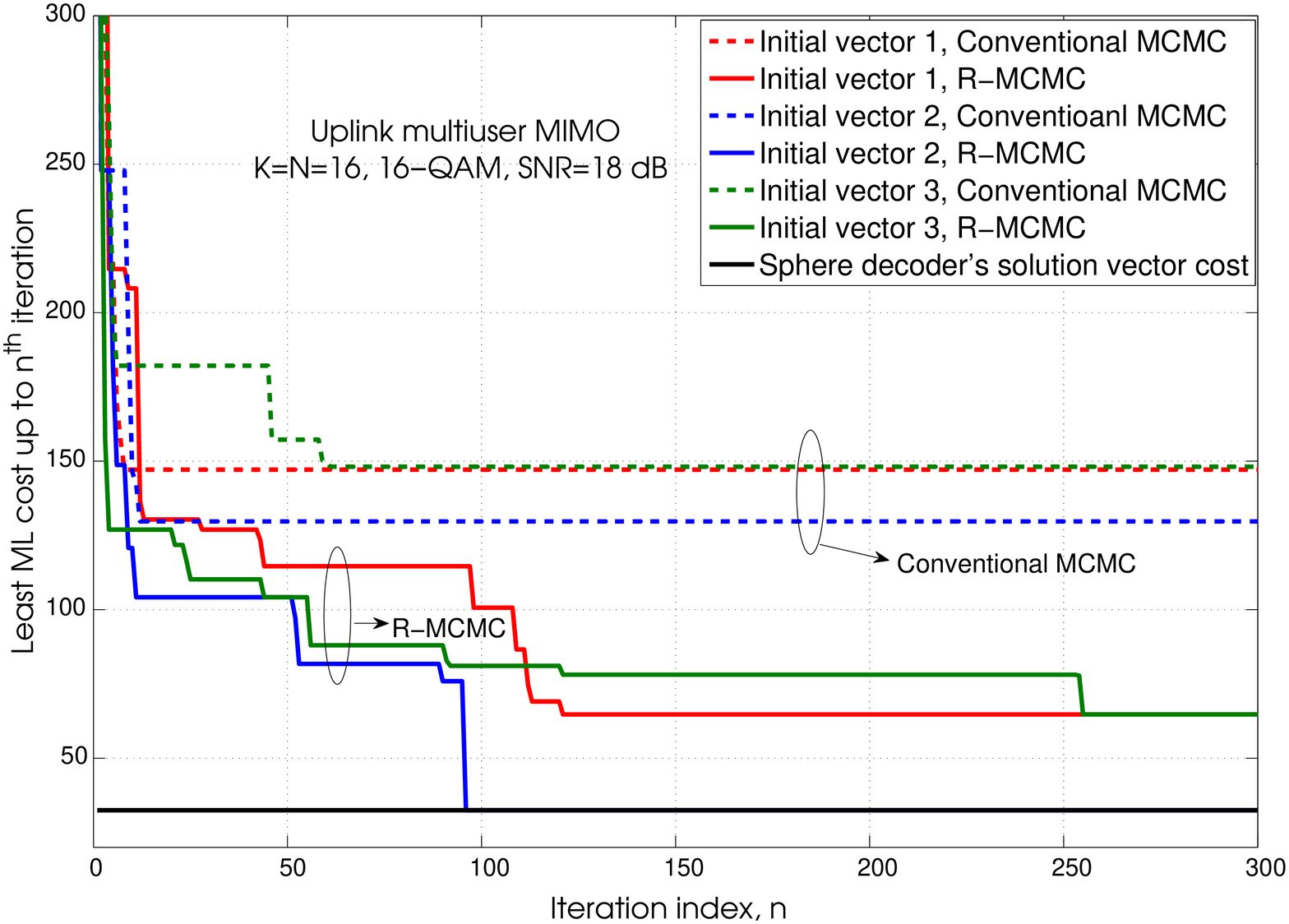}}
\vspace{-4mm}
\caption{Least ML cost up to $n$th iteration versus $n$ in conventional MCMC 
and R-MCMC for different initial vectors in multiuser MIMO with $K=N=16$.}
\vspace{-4mm}
\label{fig8}
\end{figure}

In Figs. \ref{fig8}(a) and \ref{fig8}(b), we compare the effect of multiple 
random restarts in R-MCMC and conventional MCMC algorithms for 4-QAM 
and 16-QAM, respectively. For a given realization of ${\bf x}, {\bf H}$ and
${\bf n}$, we ran both algorithms for three different random initial vectors,
and plotted the least ML cost up to $n$th iteration as a function of $n$.
We show the results of this experiment for multiuser MIMO with $K=N=16$ at
11 dB SNR for 4-QAM and 18 dB SNR for 16-QAM (these SNRs give about $10^{-3}$
BER with sphere decoding for 4-QAM and 16-QAM, respectively). The true ML
vector cost (obtained through sphere decoder simulation for the same
realization) is also plotted. It is seen that R-MCMC achieves much better 
least ML cost compared to conventional MCMC. This is because conventional 
MCMC gets locked up in some state (with very low state transition probability) 
for long time without any change in ML cost in subsequent iterations, whereas 
the randomized sampling strategy in R-MCMC is able to exit from such states 
quickly and give improved ML costs in subsequent iterations. This shows that 
R-MCMC is preferred over conventional MCMC. Even more interestingly, comparing 
the least ML costs of 4-QAM and 16-QAM (in Figs. \ref{fig8}(a) and (b), 
respectively), we see that all the three random initializations could converge 
to almost true ML vector cost for 4-QAM within 100 iterations, whereas only 
initial vector 3 converges to near true ML cost for 16-QAM and initial vectors 
1 and 2 do not. Since any random initialization works well with 4-QAM, R-MCMC 
is able to achieve near-ML performance without multiple restarts for 4-QAM. 
However, it is seen that 16-QAM performance is more sensitive to the 
initialization, which explains the poor performance of R-MCMC without 
restarts in higher-order QAM. MMSE vector can be used as an initial vector, 
but it is not a good initialization for all channel realizations. This points
to the possibility of achieving good initializations through multiple restarts 
to improve the performance of R-MCMC in higher-order QAM.

\subsubsection{R-MCMC with multiple restarts}
\label{sec3c2}
\begin{figure}
\center
\includegraphics[totalheight=9.0cm,width=12.00cm]{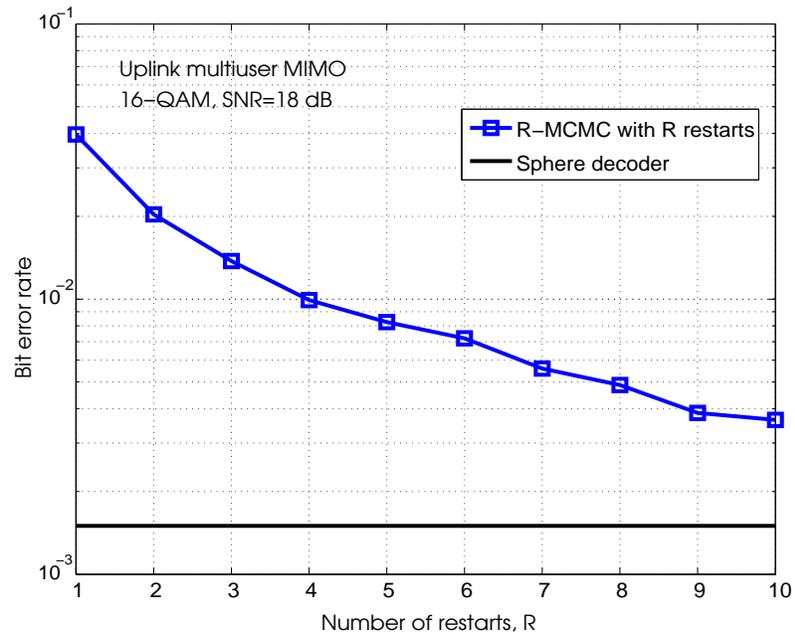}
\vspace{-6mm}
\caption{BER performance of R-MCMC as a function of number of restarts 
in multiuser MIMO with $K=N=16$ and 16-QAM at SNR = 18 dB.} 
\vspace{-4.0mm}
\label{fig9}
\end{figure}

In R-MCMC with multiple restarts, we run the basic R-MCMC algorithm multiple 
times, each time with a different random initial vector, and choose that vector 
with the least ML cost at the end as the solution vector. Figure \ref{fig9} 
shows the improvement in the BER performance of R-MCMC as the number of restarts 
($R$) is increased in multiuser MIMO with $K=N=16$ and 16-QAM at SNR = 18 dB. 
300 iterations are used in each restart. It can be observed that, though BER 
improves with increasing $R$, much gap still remains between sphere decoder
performance and R-MCMC performance even with $R=10$. A larger $R$ could get 
the R-MCMC performance close to sphere decoder performance, but at the cost 
of increased complexity. While a small $R$ results in poor performance, a 
large $R$ results in high complexity. So, instead of arbitrarily fixing $R$, 
there is a need for a good restart criterion that can significantly enhance 
the performance without incurring much increase in complexity. We devise one 
such criterion below.

\subsubsection{Proposed restart criterion}
\label{sec3c3}
\begin{figure}
\hspace{-4mm}
\subfigure[4-QAM]{
\includegraphics[width=3.5in,height=3.00in]{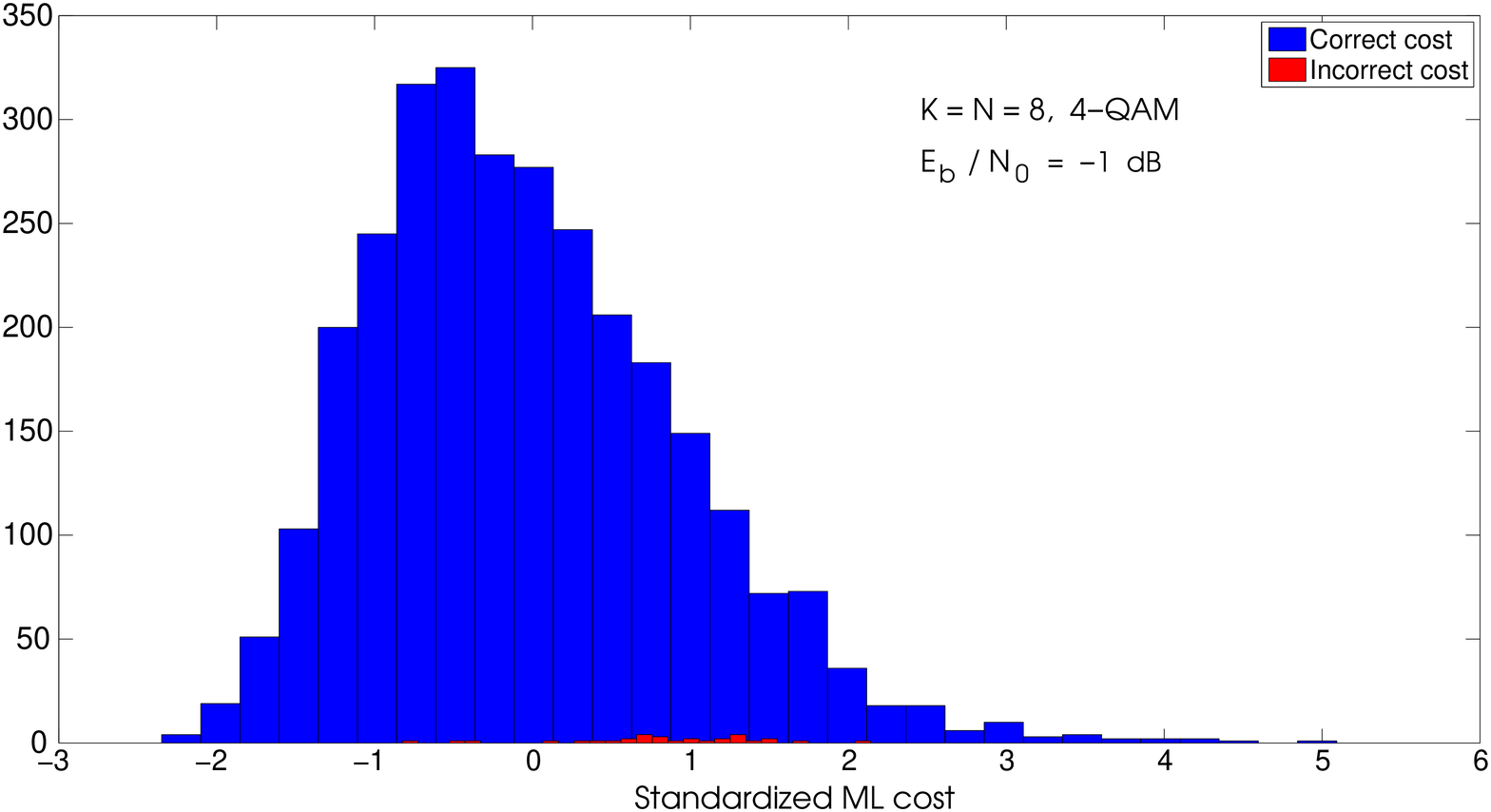}}
\hspace{-9mm}
\subfigure[16-QAM ]{
\includegraphics[width=3.5in,height=3.00in]{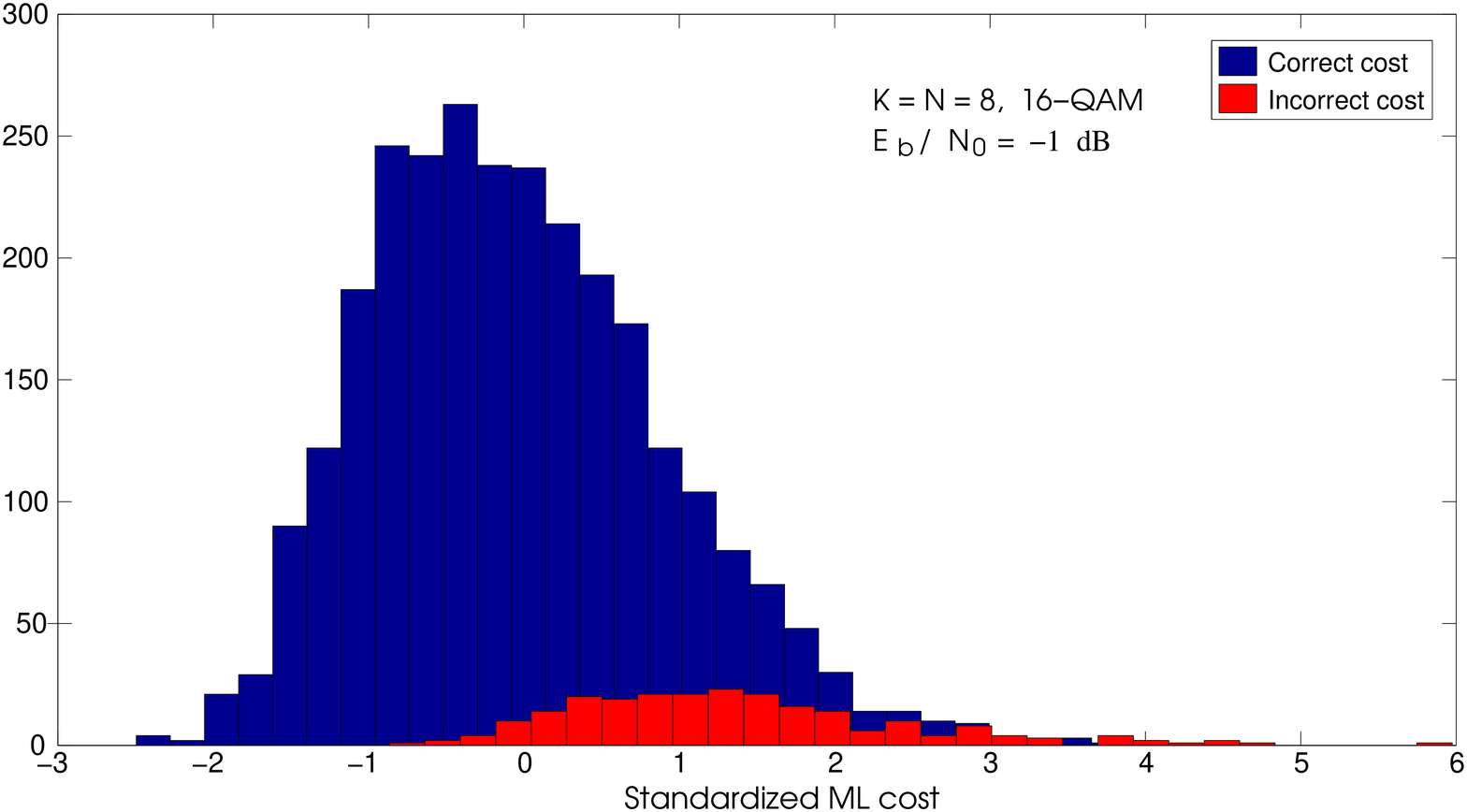}}
\vspace{-0mm}
\caption{Histograms of standardized ML costs of correct and incorrect outputs from
R-MCMC with restarts in multiuser MIMO with $K=N=8$ and 4-/16-QAM.}
\vspace{-6mm}
\label{fig10}
\end{figure}

At the end of each restart, we need to decide whether to terminate the algorithm 
or to go for another restart. To do that, we propose to use 
\begin{itemize}
\item 	the standardized ML costs (given by (\ref{qual})) of solution vectors, and 
\item 	the number of repetitions of the solution vectors.
\end{itemize}
Nearness of the ML costs obtained so far to the error-free ML cost in terms of 
its statistics can allow the algorithm to get near ML solution. Checking for 
repetitions can allow restricting the number of restarts, and hence the 
complexity. In particular, we define multiple thresholds that 
divide the range of the distribution of $\|{\bf n}\|^2$ (i.e., $\mathbb{R}^+$) 
into multiple regions, and define one integer threshold for each of these regions 
for the purpose of comparison with number of repetitions. We use the minimum 
standardized ML cost obtained so far and its number of repetitions to decide
the credibility of the solution. In Fig \ref{fig10}, we plot histograms of the
standardized ML cost of correct and incorrect solution vectors at the output
of R-MCMC with restarts in multiuser MIMO with $K=N=8$ and 4-/16-QAM. We 
judge the correctness of the obtained solution vector from R-MCMC output by 
running sphere decoder simulation for the same realizations. It can be observed 
in Fig. \ref{fig10}
that the incorrect standardized ML cost density does not stretch into negative 
values. Hence, if the obtained solution vector has negative standardized ML cost, 
then it can indeed be correct with high probability. But as the standardized ML 
cost increases in the positive domain, the reliability of that vector decreases 
and hence it would require more number of repetitions for it to be trusted as 
the final solution vector. It can also be observed from Fig. \ref{fig10} that 
the incorrect density in case of 16-QAM is much more than that of 4-QAM for the 
same SNR. So it is desired that, for a standardized ML cost in the positive 
domain, the number of repetitions needed to declare as the final solution
should increase with the QAM size. Accordingly, the number of repetitions 
needed for termination ($P$) is chosen as per the following expression: 
\begin{eqnarray}
P = \left \lfloor \max\left(0,c_2\phi(\tilde{{\bf x}})\right)\right\rfloor+1,
\label{rep}
\end{eqnarray}
where $\tilde{{\bf x}}$ is the solution vector with minimum ML cost so far.
Now, denoting $R_{max}$ to be the maximum number for restarts, the proposed 
{\em R-MCMC with restarts} algorithm (we refer to this as the R-MCMC-R algorithm) 
can be stated as follows. 
\begin{itemize}
\item   {\em {\bfseries Step 1:}} Choose an initial vector.
\item   {\em {\bfseries Step 2:}} Run the basic R-MCMC algorithm in Sec. \ref{sec3b}. 
\item   {\em {\bfseries Step 3:}} Check if $R_{max}$ number of restarts
        have been completed. If yes, go to Step 5; else go to Step 4.
\item   {\em {\bfseries Step 4:}} For the solution vector with minimum ML cost 
	obtained so far, find the required number of repetitions needed using 
	(\ref{rep}). Check if the number of repetitions of this solution vector 
	so far is less than the required number of repetitions computed in Step 4.
        If yes, go to Step 1, else go to Step 5.
\item   {\em {\bfseries Step 5:}} Output the solution vector with the minimum
	ML cost so far as the final solution. 
\end{itemize}

\subsubsection{Performance and complexity of the R-MCMC-R Algorithm}
\label{sec3c4}
\begin{figure}
\center
\includegraphics[totalheight=9.0cm,width=12.00cm]{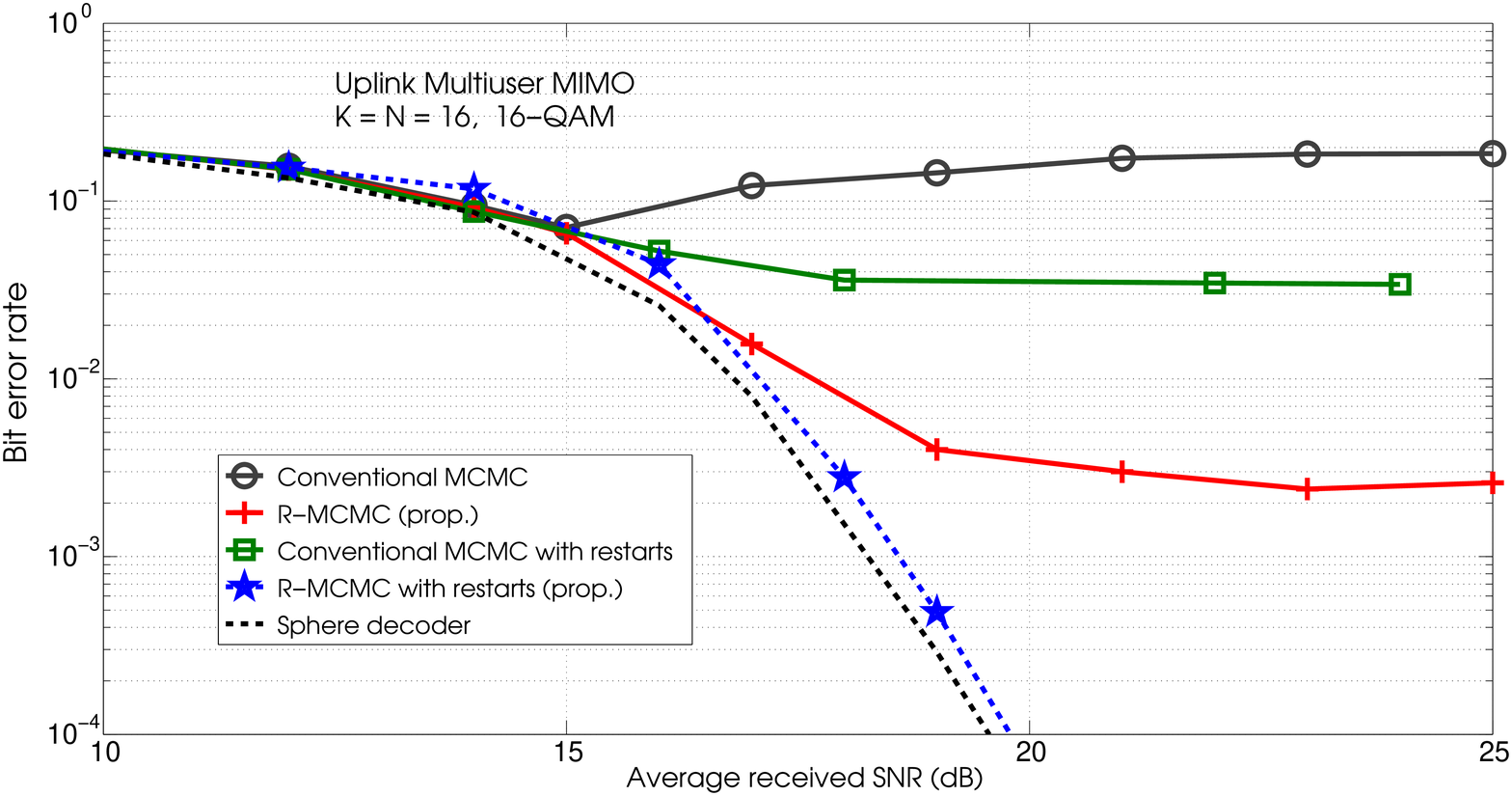}
\vspace{-6mm}
\caption{BER performance between conventional MCMC (without and with restarts), 
proposed R-MCMC (without and with restarts), and sphere decoder in uplink 
multiuser MIMO with $K=N=16$ and 16-QAM.}
\vspace{-4.0mm}
\label{fig11}
\end{figure}

The BER performance and complexity of the R-MCMC-R algorithm are evaluated 
through simulations. The following parameters are used in the simulations
of R-MCMC and R-MCMC-R:
$c_{min}=10$, $c_1=10\log_2M$ (i.e., $c_2=20,40,60$ for 4-/16-/64-QAM, 
respectively), MAX-ITER = $8K\sqrt{M}$, $R_{max}=50$, and $c_2=0.5\log_2M$.
In Fig. \ref{fig11}, we compare the BER performance of conventional MCMC, 
R-MCMC, R-MCMC-R and sphere decoder in multiuser MIMO with $K=N=16$ and
16-QAM. In the first restart, MMSE solution vector is used as the initial 
vector. In the subsequent restarts, random initial vectors are used. For 
64-QAM, the randomized sampling is applied only to the one-symbol away 
neighbors of the previous iteration index; this helps to reduce complexity 
in 64-QAM. From Fig. \ref{fig11}, it is seen that the performance of conventional 
MCMC, either without or with restarts, is quite poor. That is, using restarts 
in conventional MCMC is not of much help. This shows the persistence of the 
stalling problem. The performance of R-MCMC (without restarts) is better than 
conventional MCMC with and without restarts, but its performance still is far 
from sphere decoder performance. This shows that R-MCMC alone (without restarts) 
is inadequate the alleviate the stalling problem in higher-order QAM. However, 
the randomized sampling in R-MCMC when used along with restarts (i.e., R-MCMC-R) 
gives strikingly improved performance. In fact, the proposed R-MCMC-R algorithm 
achieves almost sphere decoder performance (close to within 0.4 dB at $10^{-3}$ 
BER). This points to the important observations that application of any one of 
the two features, namely, randomized sampling and restarts, to the conventional 
MCMC algorithm is not adequate, and that simultaneous application of both these 
features is needed to alleviate the stalling problem and achieve near-ML 
performance in higher-order QAM. 

\begin{figure}
\hspace{-4mm}
\subfigure[$K = N =16$]{
\includegraphics[width=3.5in,height=3.00in]{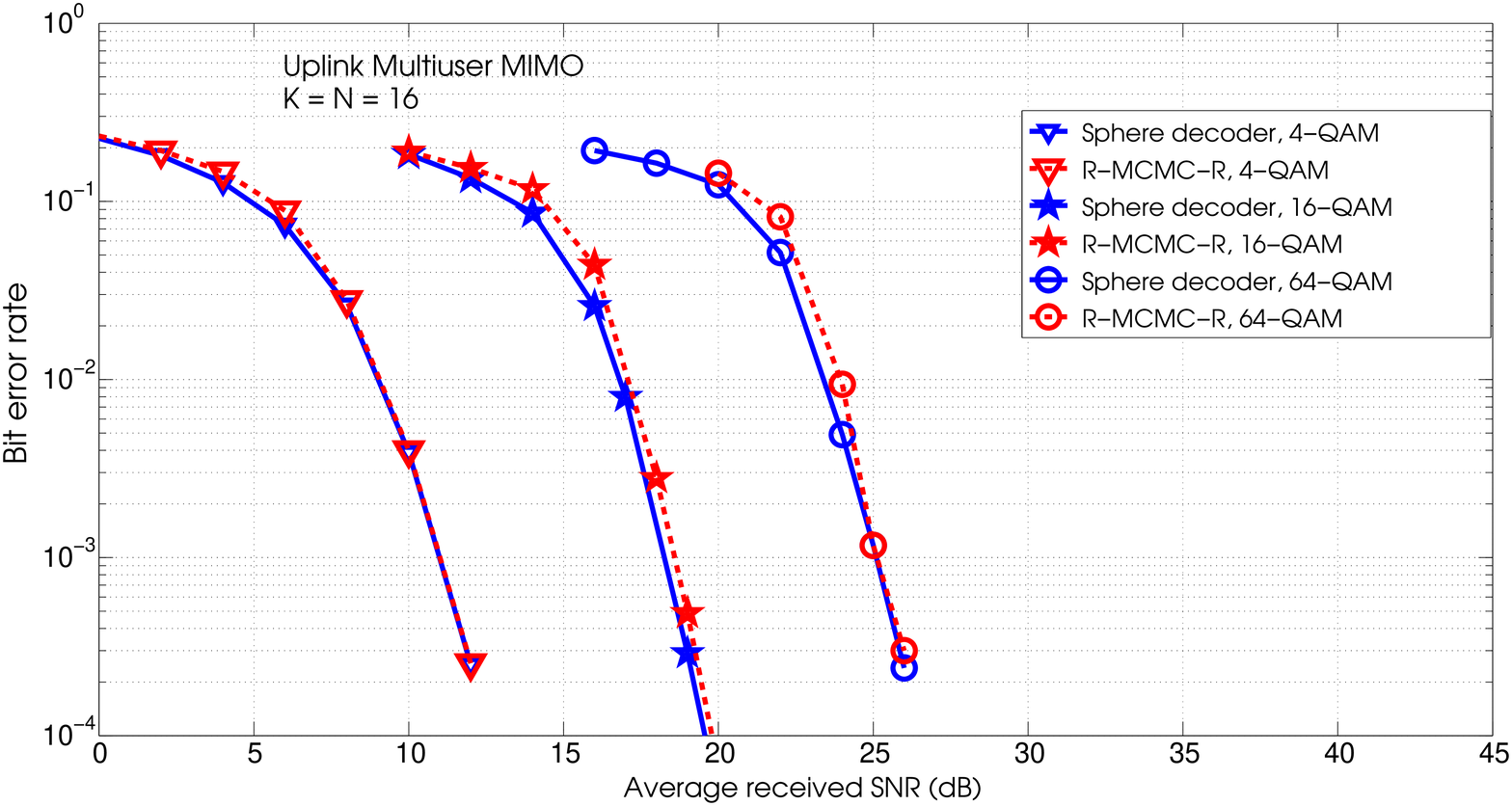}}
\hspace{-9mm}
\subfigure[$K = N = 32$]{
\includegraphics[width=3.5in,height=3.00in]{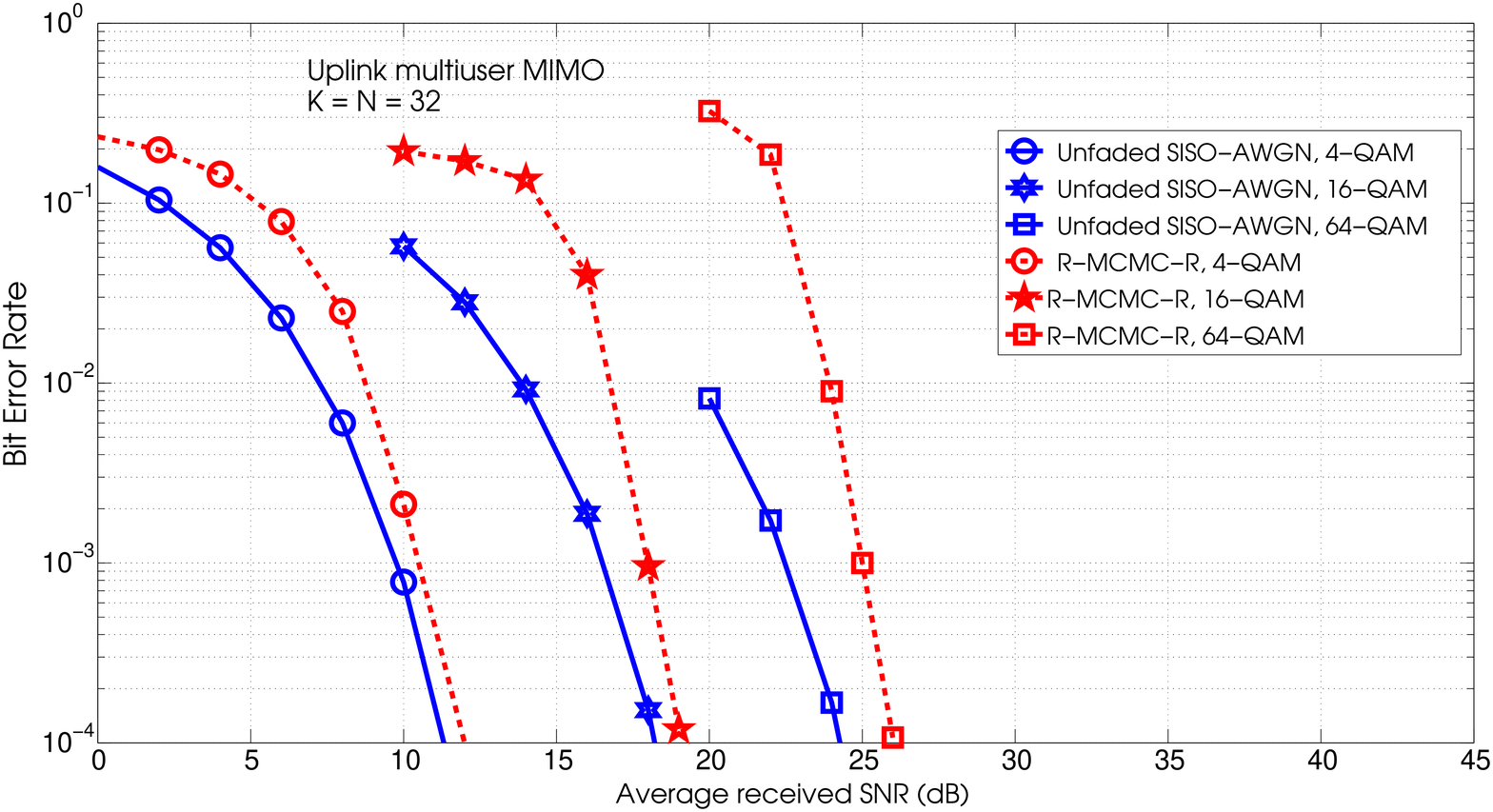}}
\vspace{-0mm}
\caption{
BER performance of R-MCMC-R algorithm in uplink multiuser MIMO with
$K=N=16$ and $32$ and higher-order modulation (4-/16-/64-QAM).} 
\vspace{-6mm}
\label{fig12}
\end{figure}

Figure \ref{fig12}(a) shows that the R-MCMC-R algorithm is able to achieve almost 
sphere decoder performance for 4-/16-/64-QAM in multiuser MIMO with $K=N=16$.
Similar performance plots for 4-/16-/64-QAM for $K=N=32$ are shown in Fig. 
\ref{fig12}(b), where the performance of R-MCMC-R algorithm is seen to be quite 
close to unfaded SISO-AWGN performance, which is a lower bound on true ML 
performance.

\subsubsection{Performance/complexity comparison with other detectors} 
\label{sec3c5}
In Table-\ref{tab1}, we present a comparison of the BER performance and 
complexity of the proposed R-MCMC-R algorithm with those of other detectors 
in the literature. Comparisons are made for systems with $K=N=16,32$ and
4-/16-/64-QAM. Detectors considered for comparison include: $i)$ 
random-restart reactive tabu search (R3TS) algorithm reported recently in 
\cite{r3ts}, which is a local neighborhood search based algorithm, and 
$ii)$ fixed-complexity sphere decoder (FSD) reported in \cite{fsd}, which 
is a sub-optimal variant of sphere decoder whose complexity is fixed 
regardless of the operating SNR. 
Table-\ref{tab1} shows the complexity measured in average number of real
operations at a BER of $10^{-2}$ and the SNR required to achieve $10^{-2}$
BER for the above three detection algorithms. It can be seen that both 
R-MCMC-R and R3TS perform better than FSD. Also, R-MCMC-R achieves the best 
performance at the lowest complexity compared to R3TS and FSD for $K=N=16$ 
with 16-QAM and 64-QAM. In 4-QAM and in $K=N=32$, R-MCMC-R achieves same or 
slightly better performance than R3TS at some increased complexity compared 
to R3TS.

\section{Proposed MCMC based channel estimation}
\label{sec4}
In the previous section, we assumed perfect channel knowledge at the BS receiver. 
In this section, we relax the perfect channel knowledge assumption and propose an 
MCMC based channel estimation algorithm.

\subsection{System model}
\label{sec4a}
\begin{figure}
\center
\includegraphics[totalheight=4.25in,width=7.5in]{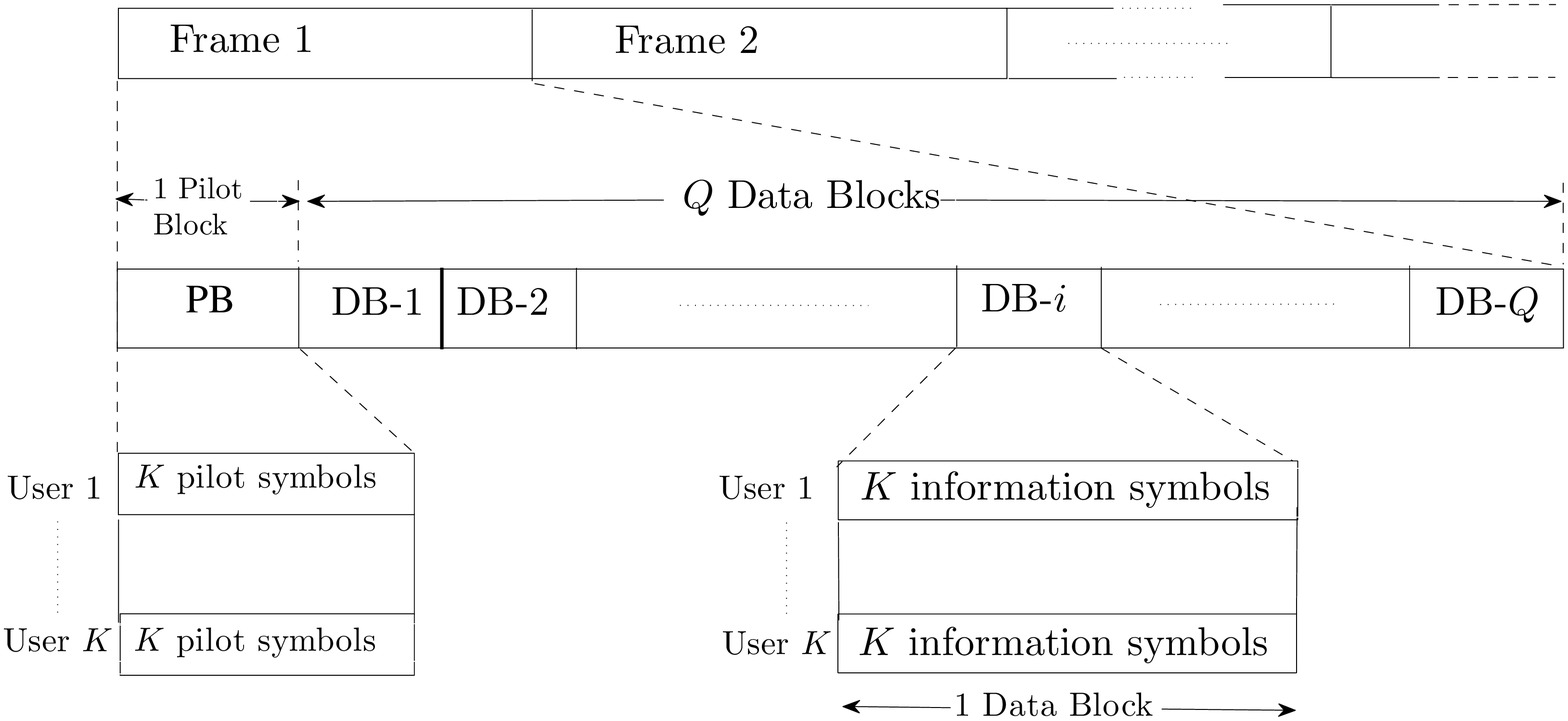}
\vspace{-30mm}
\caption{Frame structure for multiuser MIMO system in frequency 
non-selective fading.}
\label{fig13}
\end{figure}

Consider the uplink multiuser MIMO system model in (\ref{eqn1}). As in Sec. 
\ref{sec2}, perfect synchronization among users' transmissions is assumed. 
But the assumption of perfect knowledge of the channel matrix at the BS is 
relaxed here. The channel matrix is estimated based on a pilot based channel 
estimation scheme. Transmission is carried out in frames, where each frame 
consists of several blocks as shown in Fig. \ref{fig13}. A slow fading channel 
(typical with no/low mobility users) is assumed, where the channel is assumed 
to be constant over one frame duration. Each frame consists of a pilot block 
(PB) for the purpose of initial channel estimation, followed by $Q$ data blocks 
(DB). The pilot block consists of $K$ channel uses in which a $K$-length pilot 
symbol vector comprising of pilot symbols transmitted from $K$ users (one pilot 
symbol per user) is received by $N$ receive antennas at the BS. Each data block 
consists of $K$ channel uses, where $K$ number of $K$-length information symbol 
vectors (one data symbol from each user) are transmitted. Taking both pilot and 
data channel uses into account, the total number of channel uses per frame is 
$(Q+1)K$. Data blocks are detected using the R-MCMC-R algorithm using an initial 
channel estimate. The detected data blocks are iteratively used to refine the 
channel estimates during data phase using the proposed MCMC based channel 
estimation algorithm.

\subsection{Initial channel estimate during pilot phase}
\label{sec4b}
Let 
${\bf x}_{\text P}^{k}=[x_{\text P}^{k}(0),x_{\text P}^{k}(1),\cdots,x_{\text P}^{k}(K-1)]$ 
denote the the pilot symbol vector transmitted from user $k$ in $K$ channel uses 
in a frame. Let 
${\bf X}_{\text P}=[({\bf x}_{\text P}^{1})^T,({\bf x}_{\text P}^{2})^T,\cdots,({\bf x}_{\text P}^{K})^T]^T$ 
denote the $K\times K$ pilot matrix formed by the pilot symbol vectors 
transmitted by the users in the pilot phase. The received signal matrix at 
the BS, ${\bf Y}_{\text P}$, of size $N\times K$ is given by
\begin{eqnarray}
{\bf Y}_{\text P}={\bf H}_c {\bf X}_{\text P} + {\bf N}_{\text P},
\label{eqn12}
\end{eqnarray}
where ${\bf N}_{\text P}$ is the $N\times K$ noise matrix at the BS.
We use the pilot sequence given by
\begin{equation}
\label{eqn13}
\mathbf{\bf x}_{\text P}^{k}=[\mathbf{0}_{(k-1)\times 1} \quad p\quad \mathbf{0}_{(K-k)\times 1}].
\end{equation}
We choose $p=\sqrt{KE_s}$, where $E_s$ is the average symbol energy.
Using the scaled identity nature of ${\bf x}_{\text P}$, an initial channel 
estimate $\widehat{{\bf H}}_c$ is obtained as 
\begin{equation}
\label{eqn14}
\widehat{{\bf H}}_c={\bf Y}_{\text P}/p.
\end{equation}

\subsection{Data detection using initial channel estimate }
\label{sec4c}
Let ${\bf x}_i^{k}=[x_i^{k}(0),x_i^{k}(1),\cdots,x_i^{k}(K-1)]$ denote the 
data symbol vector transmitted from user $k$ in $K$ channel uses during the
$i$th data block in a frame. Let 
${\bf X}_i=[({\bf x}_i^{1})^T,({\bf x}_i^{2})^T,\cdots,({\bf x}_i^{K})^T]^T$ 
denote the $K\times K$ data matrix formed by the data symbol vectors transmitted 
by the users in the $i$th data block during data phase, $i=1,2,\cdots,Q$. The 
received signal matrix at the BS in the $i$th data block, ${\bf Y}_i$ of size 
$N\times K$, is given by
\begin{eqnarray}
{\bf Y}_i={\bf H}_c {\bf X}_i + {\bf N}_i,
\label{eqn15}
\end{eqnarray}
where ${\bf N}_i$ is the $N\times K$ noise matrix at the BS during $i$th data 
block. We perform the detection on a vector by vector basis using the independence 
of data symbols transmitted by the users. Let ${\bf y}_i^{(t)}$ denote the $t$th 
column of ${\bf Y}_i$, $t=0,2,\cdots,K-1$. Denoting the $t$th column of ${\bf X}_i$ 
as ${\bf x}_i^{(t)}=[x_i^{1}(t),x_i^{2}(t),\cdots,x_i^{K}(t)]^T$, we can rewrite 
the system equation (\ref{eqn12}) as
\begin{eqnarray}
{\bf y}_i^{(t)}={\bf H}_c {\bf x}_i^{(t)} + {\bf n}_i^{(t)},
\label{eqn16}
\end{eqnarray}
where ${\bf n}_i^{(t)}$ is the $t$th column of ${\bf N}_i$.
The initial channel estimate $\widehat{{\bf H}}_c$ obtained from 
(\ref{eqn14}) is used to detect the transmitted data vectors using 
the R-MCMC-R algorithm presented in Sec. \ref{sec3}. 

From (\ref{eqn12}) and (\ref{eqn14}), we observe that  
$\widehat{{\bf H}}_c= {\bf H}_c+{\bf N}_{\text P}/p$. This knowledge about 
imperfection of channel estimates is used to calculate the statistics of 
error-free ML cost required in the R-MCMC-R algorithm. In Sec. \ref{sec3}, we 
have observed that in case of perfect channel knowledge, the error-free ML cost 
is nothing but $\|{\bf n}^2\|$. In case of imperfect channel knowledge at the
receiver, at channel use $t$,
\begin{eqnarray*}
\|{\bf y}_i^{(t)}-\widehat{{\bf H}}_c{\bf x}_i^{(t)}\|^2=\|{\bf n}_i^{(t)}-{\bf N}_{\text P}{\bf x}_i^{(t)}/p\|^2.
\end{eqnarray*}
Each entry of the vector ${\bf n}_i^{(t)}- {\bf N}_{\text P}{\bf x}_i^{(t)}/p$ 
has mean zero and variance $2\sigma^2$. Using this knowledge at the receiver, 
we detect the transmitted data using R-MCMC-R algorithm and obtain 
${\bf \widehat{x}}_i^{(t)}$. Let the detected data matrix in data block $i$ be 
denoted as 
${\bf \widehat{ X}}_i$ = $[{\bf \widehat{x}}_i^{(0)}$, ${\bf \widehat{x}}_i^{(1)}$, $\cdots$, ${\bf \widehat{x}}_i^{(K-1)}]$.

\subsection{Channel estimation using MCMC algorithm in data phase}
\label{sec4d}
Let 
${\bf Y}_{tot}=[{\bf Y}_{\text P}\, {\bf Y}_1\, {\bf Y}_2\, \cdots\, {\bf Y}_Q]$, 
${\bf X}_{tot}=[{\bf X}_{\text P}\, {\bf X}_1\, {\bf X}_2\, \cdots\, {\bf X}_Q]$, 
and 
${\bf N}_{tot}=[{\bf N}_{\text P}\, {\bf N}_1\, {\bf N}_2\, \cdots\, {\bf N}_Q]$ 
denote the matrices corresponding to one full frame.
We can express ${\bf Y}_{tot}$ as
\begin{eqnarray}
 {\bf Y}_{tot}={\bf H}_c {\bf X}_{tot} +{\bf N}_{tot}.
\label{eqn17}
\end{eqnarray}
This system model corresponding to the full frame is converted into a real-valued 
system model as done in Sec. \ref{sec2}. That is, (\ref{eqn17}) can be written 
in the form 
\begin{eqnarray}
 {\bf Y}={\bf H}{\bf X} +{\bf N},
\label{eqn18}
\end{eqnarray}
where
\[
{\bf Y}=\left[\begin{array}{cc}\Re({\bf Y}_{tot}) \hspace{2mm} -\Im({\bf Y}_{tot}) \\
\Im({\bf Y}_{tot}) \hspace{5mm} \Re({\bf Y}_{tot}) \end{array}\right], \quad
{\bf H} = \left[\begin{array}{cc}\Re({\bf H}_c) \hspace{2mm} -\Im({\bf H}_c) \\
\Im({\bf H}_c) \hspace{5mm} \Re({\bf H}_c) \end{array}\right],
\]
\[
{\bf X}=\left[\begin{array}{cc}\Re({\bf X}_{tot}) \hspace{2mm} -\Im({\bf X}_{tot}) \\
\Im({\bf X}_{tot}) \hspace{5mm} \Re({\bf X}_{tot}) \end{array}\right], \quad
{\bf N}=\left[\begin{array}{cc}\Re({\bf N}_{tot}) \hspace{2mm} -\Im({\bf N}_{tot}) \\
\Im({\bf N}_{tot}) \hspace{5mm} \Re({\bf N}_{tot}) \end{array}\right].
\]
Equation (\ref{eqn18}) can be written as 
\begin{equation}
{\bf Y}^{T} = {\bf X}^{T}{\bf H}^{T} + {\bf N}^{T}. 
\label{eqn18a}
\end{equation}
Vectorizing the matrices ${\bf Y}^{T}$, ${\bf H}^{T}$, and ${\bf N}^{T}$,
we define
\begin{equation*}
{\bf r} \Define  vec({\bf Y}^T), \quad
{\bf g} \Define  vec({\bf H}^T), \quad
{\bf z} \Define  vec({\bf N}^T). \quad
\end{equation*}
With the above definitions, (\ref{eqn18a}) can be written in vector form as
\begin{eqnarray}
{\bf r}=  \underbrace{{\mathbf I}_{2N} \otimes {\bf X}^{T}}_{\Define \, {\mathbf S} } \, {\bf g}+{\bf z}.
\label{eqn20}
\end{eqnarray}
Now, our goal is to estimate ${\bf g}$ knowing ${\bf r}$, estimate of 
${\mathbf S}$ and the statistics of ${\bf z}$ using an MCMC approach. 
The estimate of ${\mathbf S}$ is obtained as 
\[
\widehat{{\bf S}} = {\mathbf I}_{2N} \otimes {\bf \widehat{X}}^{T}, 
\] 
where
$\widehat{{\bf X}}=\left[\begin{array}{cc}\Re(\widehat{{\bf X}}_{tot}) \hspace{2mm} -\Im(\widehat{{\bf X}}_{tot}) \\
\Im(\widehat{{\bf X}}_{tot}) \hspace{5mm} \Re(\widehat{{\bf X}}_{tot}) \end{array}\right]$ and 
${\bf \widehat{X}}_{tot}=[{\bf X}_{\text P}\, {\bf \widehat{X}}_1\, {\bf \widehat{X}}_2\, \cdots\, {\bf \widehat{X}}_Q]$. 
The initial vector for the MCMC algorithm is obtained as 
\begin{equation}
\widehat{{\bf g}}^{(0)} =  vec(\widehat{{\bf H}}^T),
\label{initg}
\end{equation}
where
\begin{equation}
\widehat{{\bf H}} = \left[\begin{array}{cc}\Re(\widehat{{\bf H}}_c) \hspace{2mm} -\Im(\widehat{{\bf H}}_c) \\
\Im(\widehat{{\bf H}}_c) \hspace{5mm} \Re(\widehat{{\bf H}}_c) \end{array}\right]. 
\end{equation}

\subsubsection{Gibbs sampling based estimation}
\label{sec4d1}
The vector ${\bf g}$ is of length $4KN\times 1$. To estimate ${\bf g}$, the 
algorithm starts with an initial estimate, takes samples from the conditional 
distribution of each coordinate in ${\bf g}$, and updates the estimate. This 
is carried out for a certain number of iterations. At the end of the iterations, 
a weighted average of the previous and current estimates are given as the output.

Let the $i$th coordinate in ${\bf g}$ be denoted by $g_i$, and let ${\bf g}_{-i}$ 
denote all elements in ${\bf g}$ other than the $i$th element. Let 
$\widehat{\bf s}_q$ denote the $q$th column of $\widehat{{\bf S}}$. The
conditional probability distribution for the $i$th coordinate is given by
\begin{eqnarray}
p\left(g_i|{\bf r}, {\bf \widehat{S}}, {\bf g}_{-i}\right) &\propto&   p(g_i).\,p\left({\bf r}|g_i, {\bf \widehat{S}}, {\bf g}_{-i}\right)\\
&\propto& \exp\left(-|g_i|^2\right) \exp\left(-\frac{\|{\bf r}-\sum_{q=1, q\not = i}^{4KN} g_q{\bf \widehat{s}}_q - g_i {\bf \widehat{s}}_i \|^2}{\sigma^2}\right) \\
&=& \exp\left(-|g_i|^2-\frac{\|{\widetilde{\bf r}}^{(i)}-g_i{\bf \widehat{s}}_i \|^2}{\sigma^2}\right)\\
&=& \exp\left(-\frac{\|\bar{{\bf r}}^{(i)}-g_i{\bf \bar{s}}_i \|^2}{\sigma^2}\right),
\label{eqn25a}
\end{eqnarray}
where 
${\widetilde{\bf r}}^{(i)}={\bf r}-\sum_{q=1, q\not = i}^{4KN} g_q{\bf \widehat{s}}_q $, 
$\bar{{\bf r}}^{(i)}=[{\widetilde{\bf r}}^{(i)},\, 0]^T$,
and
${\bf \bar{s}}_i=[{\bf \widehat{s}}_i ,\,\sigma]^T$.
The quantity 
$\|\bar{{\bf r}}^{(i)}-g_i{\bf \bar{s}}_i \|^2$ in (\ref{eqn25a}) is minimized for 
$g_i=\frac{\left(\bar{{\bf r}}^{(i)}\right)^T{\bf \bar{s}}_i}{\|\bar{{\bf s}}_i\|^2}$.
Hence, we can write
\begin{eqnarray}
\|\bar{{\bf r}}^{(i)}-g_i{\bf \bar{s}}_i \|^2&=&\|\bar{{\bf r}}^{(i)}-\left(\frac{ \left( \bar{{\bf r}}^{(i)}\right)^T {\bf \bar{s}}_i }{ \|\bar{{\bf s}}_i\|^2 }+g_i-\frac{ \left( \bar{{\bf r}}^{(i)}\right)^T {\bf \bar{s}}_i }{ \|\bar{{\bf s}}_i\|^2 }\right){\bf \bar{s}}_i \|^2 \nonumber\\
&=&\|\bar{{\bf r}}^{(i)}-\frac{ \left( \bar{{\bf r}}^{(i)}\right)^T {\bf \bar{s}}_i }{ \|\bar{{\bf s}}_i\|^2 }{\bf \bar{s}}_i \|^2+\left(g_i-\frac{ \left( \bar{{\bf r}}^{(i)}\right)^T {\bf \bar{s}}_i }{ \|\bar{{\bf s}}_i\|^2 }\right)^2\|{\bf \bar{s}}_i\|^2.
\end{eqnarray}
Hence, 
\begin{eqnarray}
p\left(g_i|{\bf r}, {\bf \widehat{S}}, {\bf g}_{-i}\right) &\propto&\exp\left(-\frac{\Big(g_i-\frac{ \left( \bar{{\bf r}}^{(i)}\right)^T {\bf \bar{s}}_i }{ \|\bar{{\bf s}}_i\|^2 }\Big)^2}{\frac{\sigma^2}{\|{\bf s}_i\|^2}}\right).
\end{eqnarray}
This distribution is Gaussian with mean
\begin{eqnarray}
\mu_{g_i}=\frac{\left(\bar{{\bf r}}^{(i)}\right)^T{\bf \bar{s}}_i}{\|\bar{{\bf s}}_i\|^2},
\label{eqn24}
\end{eqnarray}
and variance
\begin{eqnarray}
\sigma_{g_i}^2=\frac{\sigma^2}{2\|{\bf s}_i\|^2}.
\label{eqn25}
\end{eqnarray}
Let $MAX$ denote the number of MCMC iterations. In each MCMC iteration, for each 
coordinate, the probability distribution specified by its mean and variance 
has to be calculated to draw samples. Let the mean and variance in $r$th MCMC 
iteration and $i$th coordinate be denoted as $\mu_{g_i}^{(r)}$ and 
${\sigma_{g_i}^2}^{(r)}$, respectively, where $r=1,2,\cdots,MAX$ 
and $i=1,2,\cdots,4KN$.
We use $\widehat{{\bf g}}^{(0)} $in (\ref{initg}), which is the estimate from 
the pilot phase, as the initial estimate. In the $r$th MCMC iteration,
we obtain $\widehat{{\bf g}}^{(r)}$ from $\widehat{{\bf g}}^{(r-1)}$ as follows:
\begin{itemize}
\item 	Take $\widehat{{\bf g}}^{(r)}=\widehat{{\bf g}}^{(r-1)}$.
\item 	Update the $i$th coordinate of $\widehat{{\bf g}}^{(r)}$ 
      	by sampling from $\mathcal N\Big(\mu_{g_i}^{(r)}$, 
	${\sigma_{g_i}^2}^{(r)}\Big)$ for all $i$. 
      	Let $\widehat{g}^{(r)}_i$ denote the updated $i$th coordinate of 
      	${\bf g}^{(r)}$.  
\item 	Compute weights 
	$\alpha_i^{(r)}=\exp\left(-\frac{\left(\widehat{g}^{(r)}_i-\mu_{g_i}^{(r)}\right)^2}{2{\sigma_{g_i}^2}^{(r)}}\right)$ for all $i$. 
	This gives more weight to samples closer to the mean.
\end{itemize}
After $MAX$ iterations, we compute the final estimate of the $i$th coordinate,
denoted by $g_i^*$, to be the following weighted sum of the estimates from 
previous and current iterations:
\begin{eqnarray}
g_i^*=\frac{\sum_{r=1}^{MAX}\alpha_i^{(r)}\widehat{g}^{(r)}_i}{\sum_{r=1}^{MAX}\alpha_i^{(r)}}.
\end{eqnarray}
Finally, the updated $2N\times 2K$ channel estimate ${\widehat{{\bf H}}}$ 
is obtained by restructuring ${\bf g}^*=[g_1^*, g_2^*, \cdots, g_{4KN}^*]^T$
as follows:
\begin{eqnarray}
{\widehat{{\bf H}}}(p,q)=g_n^*, \quad p=1,2,\cdots,2N, \quad q=1,2,\cdots,2K,
\end{eqnarray}
where $n=2N(p-1)+q$ and $\widehat{{\bf H}}(p,q)$ denotes the element in the $p$th 
row and $q$th column of $\widehat{{\bf H}}$. A complete listing of the proposed 
MCMC algorithm for channel estimation is given in the next page.

The matrix $\widehat{{\bf H}}$ obtained thus is used for data detection using 
R-MCMC-R algorithm. This ends one iteration between channel estimation and 
detection. The detected data matrix is fed back for channel estimation in the 
next iteration, whose output is then used to detect the data matrix again. This 
iterative channel estimation and detection procedure is carried out for a 
certain number of iterations. 

\begin{algorithm}
\caption{Proposed MCMC algorithm for channel estimation}
\begin{algorithmic}[1]
\STATE {{\bf input:} ${\bf r}$, ,
$\widehat{{\mathbf S}}$, $\sigma^2$,
$\widehat{{\bf g}}^{(0)}:$ initial vector $\in {\mathbb{R}}^{4KN}$; \,\,
$MAX$: max. \# iterations;}
\vspace{-2mm}
\STATE $r=1$; \,\, $g^*(0)=\widehat{{\bf g}}^{(0)}$; \,\, $\alpha_i^{(0)}=0$, $\forall i=1,2,\cdots,4KN$;\\
\vspace{-2mm}
\WHILE{$r < MAX$}
\vspace{-2mm}
\STATE{ $\widehat{{\bf g}}^{(r)}=\widehat{{\bf g}}^{(r-1)}$};
\vspace{-2mm}
\STATE{${\widetilde{\bf r}}^*={\bf r}- \widehat{{\mathbf S}}\widehat{{\bf g}}^{(r)} $};
\vspace{-2mm}
\FOR{$i=1$ to $4KN$}
\vspace{-2mm}
\STATE Compute ${\widetilde{\bf r}}^{(i)}={\widetilde{\bf r}}^*+ \widehat{g}_i^{(r)}{\bf \widehat{s}}_i $, 
$\bar{{\bf r}}^{(i)}=[{\widetilde{\bf r}}^{(i)},\, 0]^T$,
and
${\bf \bar{s}}_i=[{\bf \widehat{s}}_i ,\,\sigma]^T$;\\
\STATE Compute $\mu_{g_i}^{(r)}= \frac{ \left( \bar{{\bf r}}^{(i)}\right)^T {\bf \bar{s}}_i }{ \|\bar{{\bf s}}_i\|^2 }$
 and ${\sigma_{g_i}^2}^{(r)}=\frac{\sigma^2}{2\|{\bf s}_i\|^2}$;\\
\STATE Sample $\widehat{g}_i^{(r)} \thicksim \mathcal N\Big(\mu_{g_i}^{(r)}$, ${\sigma_{g_i}^2}^{(r)}\Big)$;\\
\vspace{-2mm}
\STATE ${\widetilde{\bf r}}^*={\widetilde{\bf r}}^{(i)}- \widehat{g}_i^{(r)}{\bf \widehat{s}}_i $;\\
\vspace{-2mm}
\STATE Compute $\alpha_i^{(r)}=\exp\left(-\frac{\left(\widehat{g}^{(r)}_i-\mu_{g_i}^{(r)}\right)^2}{2{\sigma_{g_i}^2}^{(r)}}\right)$;\\
\STATE $g_i^*(r)=\frac{\alpha_i^{(r)}\widehat{g}^{(r)}_i+\left(\sum_{z=o}^{r-1}\alpha_i^{(z)}\right)g_i^*(r-1)}{\sum_{z=0}^{r}\alpha_i^{(z)}}$;\\
\vspace{-2mm}
\ENDFOR\\
\vspace{-2mm}
\STATE $r=r+1$;
\vspace{-2mm}
\ENDWHILE\\
\vspace{-2mm}
\STATE {{\bf output:} ${\bf g}^*={\bf g}^*(MAX). \qquad{\bf g}^*:$ output solution vector}\\
\end{algorithmic}
\end{algorithm}

\subsection{Performance Results}
\label{sec4f}
\begin{figure}
\hspace{-4mm}
\subfigure[MSE]{
\includegraphics[width=3.5in,height=3.00in]{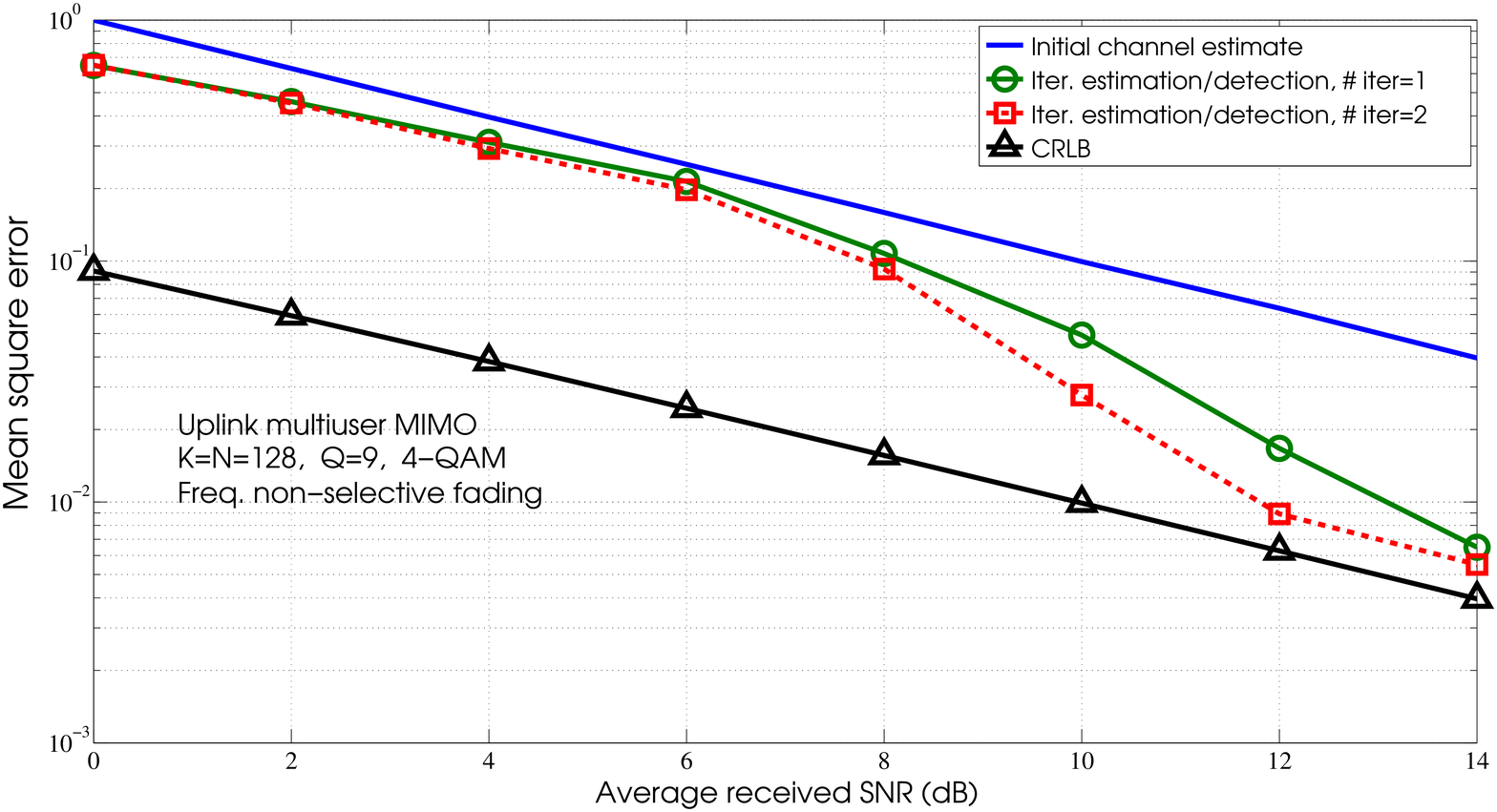}}
\hspace{-9mm}
\subfigure[BER]{
\includegraphics[width=3.5in,height=3.00in]{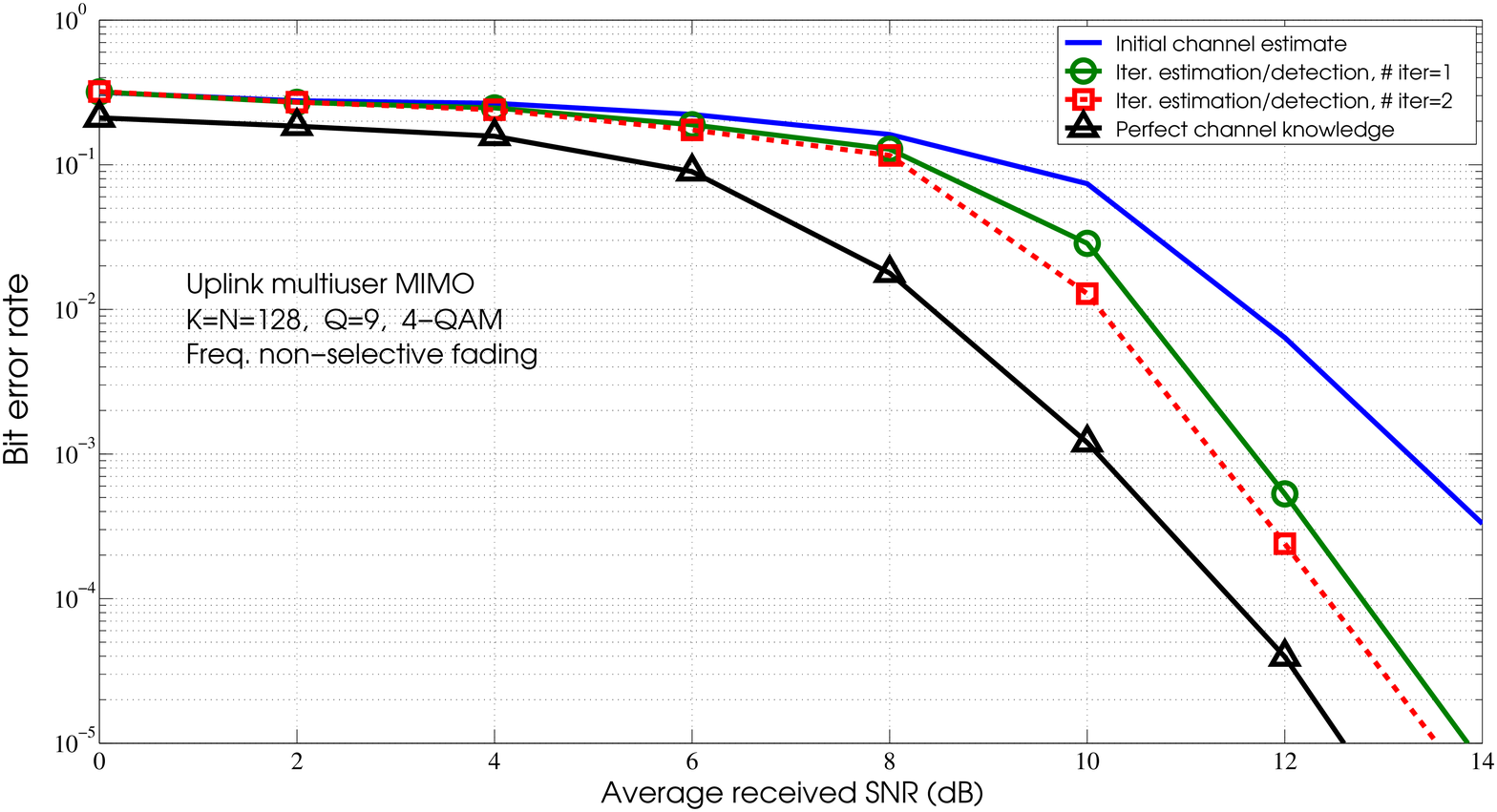}}
\vspace{-0mm}
\caption{
MSE and BER performance of iterative channel estimation/detection using MCMC 
based channel estimation and R-MCMC-R based detection in uplink multiuser MIMO 
systems on frequency non-selective fading with $K=N=128, Q=9$, 4-QAM.}
\vspace{-6mm}
\label{fig14}
\end{figure}

In Fig. \ref{fig14}(a), we plot the mean square error performance (MSE) of the 
iterative channel estimation/detection scheme using proposed MCMC based channel 
estimation and R-MCMC-R based detection with 4-QAM for $K=N=128$ and $Q=9$. In 
the simulations, the R-MCMC-R algorithm parameter values used are the same as 
in Sec. \ref{sec3c4}. For the MCMC channel estimation algorithm, the value of 
MAX used is 2. The MSEs of the initial channel estimate, and the channel estimates 
after 1 and 2 iterations between channel estimation and detection are shown. For 
comparison, we also plot the Cramer-Rao lower bound (CRLB) for this system. It 
can be seen that in the proposed scheme results in good MSE performance with 
improved MSE for increased number of iterations between channel estimation and 
detection. For the same set of system and algorithm parameters in Fig. 
\ref{fig14}(a), we plot the BER performance curves in Fig. \ref{fig14}(b). For 
comparison, we also plot the BER performance with perfect channel knowledge. 
It can be seen that with two iterations between channel estimation and detection 
the proposed MCMC based algorithms can achieve $10^{-3}$ BER within about 1 dB 
of the performance with perfect channel knowledge.

\section{Equalization/Channel Estimation in Frequency Selective Fading}
\label{sec5}
In the previous sections, we considered frequency non-selective fading. In this 
section, we consider equalization and channel estimation in frequency selective 
fading. Multicarrier techniques like OFDM can transform a frequency selective 
channel into several narrow-band frequency non-selective channels. The MCMC based
algorithms proposed in the previous sections can be employed for signal detection 
and channel estimation on the resulting narrow-band channels. However, multicarrier 
systems face the peak-to-average power ratio (PAPR) problem. Single carrier (SC) 
block transmission schemes are considered as good alternatives to address the PAPR 
issue that arises in multicarrier systems \cite{sc1}-\cite{sc7}. We consider 
cyclic prefixed single-carrier (CPSC) signaling, where the overall channel 
includes a Fourier transform (FFT) operation so that the transmitted symbols 
are estimated from the received frequency-domain signal 
\cite{sc1},\cite{sc2},\cite{sc6}. In \cite{opt_train}, the optimal training 
sequence that minimizes the channel estimation mean square error of the 
linear channel estimator is shown to be of length $KL$ per transmit antenna. 
Blind/semi-blind channel estimation methods can be considered, but they 
require long data samples and the complexity is high 
\cite{chl_est_zp2},\cite{chl_est_zp3}. Here, we consider channel estimation 
using uplink pilots and iterations between channel estimation and equalization
in multiuser MIMO CPSC systems.

\subsection{Multiuser MIMO CPSC system model}
\label{sec5a}
\begin{figure}
\center
\includegraphics[totalheight=4.25in,width=7.5in]{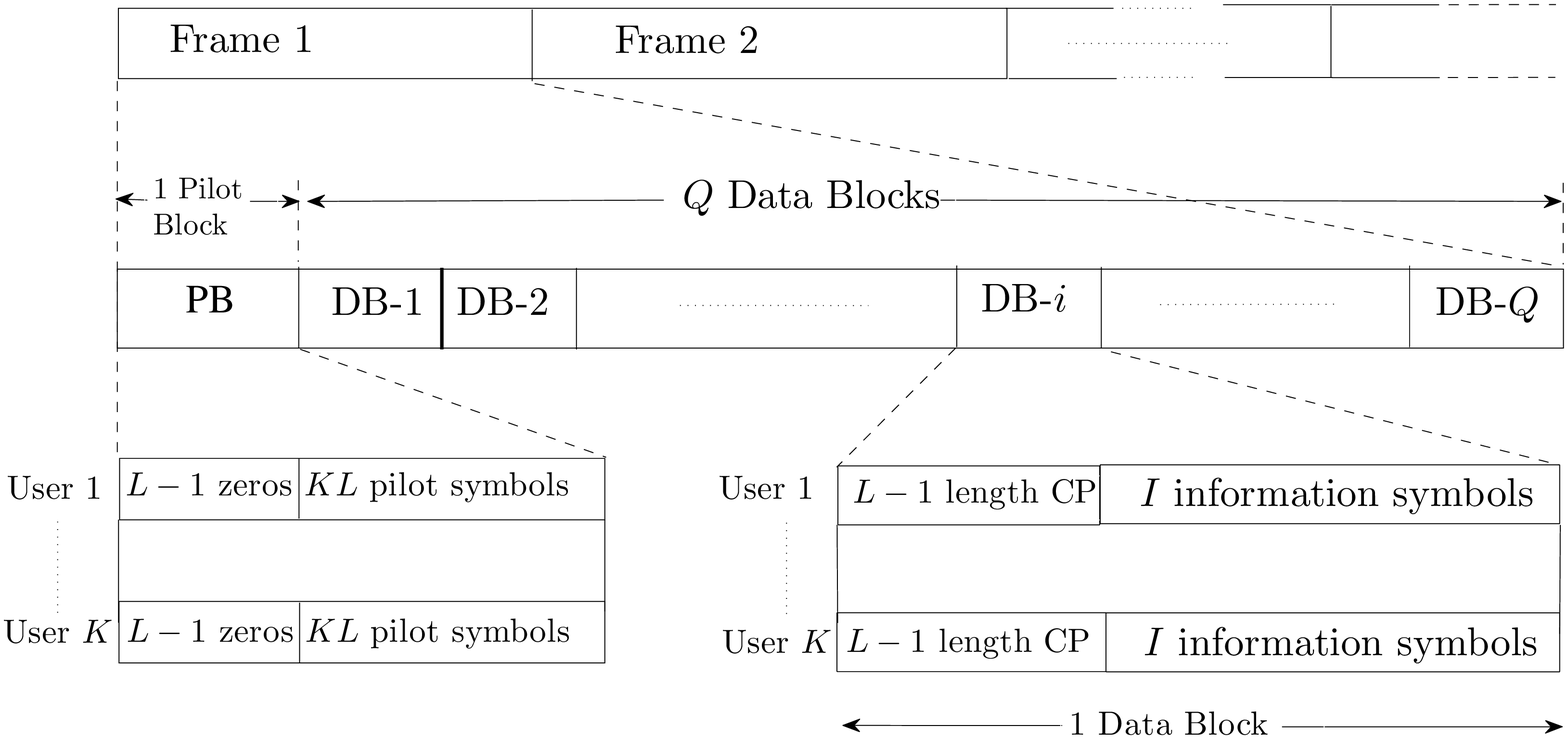}
\vspace{-28mm}
\caption{Frame structure for multiuser MIMO CPSC system in frequency selective 
fading.}
\label{fig15}
\vspace{-4.0mm}
\end{figure}

Consider the uplink multiuser MIMO system shown in Fig. \ref{fig1}. The channel
between each pair of user transmit antenna and BS receive antenna is assumed to 
be frequency selective with \textit{L} multipath components. Let $h^{(j,k)}(l)$
denote the channel gain between $k$th user and $j$th receive antenna at the BS 
on the $l$th path, which is modeled as $\mathcal{CN}(0,\Omega _{l}^2)$. As in 
Sec. \ref{sec2}, perfect synchronization among users' transmissions is assumed.
Transmission is carried out in frames, where each frame consists of several 
blocks as shown in Fig. \ref{fig15}. As in \ref{sec4a}, the channel is assumed 
to be constant over one frame duration. Each frame consists of a pilot block 
for the purpose of initial channel estimation, followed by $Q$ data blocks. 
The pilot block consists of $(L-1)+KL$ channel uses. In the first $L-1$ channel 
uses in the pilot block, padding of $L-1$ zeros is used to avoid inter-frame 
interference. In each of the remaining $KL$ channel uses, a $K$-length pilot 
symbol vector comprising of pilot symbols transmitted from $K$ users (one pilot 
symbol per user) is received by $N$ receive antennas at the BS. Each data block 
consists of $I+L-1$ channel uses, where $I$ number of $K$-length information 
symbol vectors (one data symbol from each user) preceded by $(L-1)$-length 
cyclic prefix from each user (to avoid inter-block interference) are transmitted. 
With $Q$ data blocks in a frame, the number of channel uses in the data part of 
the frame is $(I+L-1)Q$. Taking both pilot and data channel uses into account, 
the total number of channel uses per frame is $(L+1)K+(I+L-1)Q-1$. Data blocks 
are detected using the R-MCMC-R algorithm using an initial channel estimate. 
The detected data blocks are then iteratively used to refine the channel 
estimates during data phase. The padding of $L-1$ zeros at the beginning of 
the pilot block makes the transmitters silent during the first $L-1$ channel 
uses in a frame. The channel output in these channel uses are ignored at the 
receiver. Accordingly, the 0th channel use in a frame at the receiver is taken 
to be the channel use in which the first pilot symbol in the frame is sent.

\subsection{Initial channel estimate during pilot phase}
\label{sec5b}
Let $\mathbf{b}^{k}=[b^{k}(0),b^{k}(1),\cdots,b^{k}(KL-1)]$ denote the pilot 
symbol vector transmitted from user $k$ in $KL$ channel uses in a frame. 
The signal received by the $j$th receive antenna at the BS during pilot phase 
in the $n$th channel use is given by
\begin{equation}
\label{eq2}
y_{\text P}^{j}(n)=\sum_{k=1}^{K}\sum_{l=0}^{L-1}h^{(j,k)}(l)b^{k}(n-l)+{q}_{\text P}^{j}(n),
\end{equation}
$j=1,2\cdots,N$, $n=0,1,\cdots,KL-1$, where the subscript ${\text P}$
in ${y}_{\text P}^{j}(n)$ and ${q}_{\text P}^{j}(n)$ denotes pilot phase.
$\{{q}_{\text P}^{j}(n)\}$ are noise samples modeled as i.i.d.
$\mathcal{CN}(0,\sigma^2)$.
We use the training sequence given by
\begin{equation}
\label{eq1}
\mathbf{b}^{k}=[\mathbf{0}_{(k-1)L\times 1} \quad b\quad \mathbf{0}_{(K-(k-1))L-1)\times 1}].
\end{equation}
Writing (\ref{eq2}) in matrix notation after substituting (\ref{eq1}), we get
\begin{equation}
\label{eq3}
\mathbf{y}_{\text P}^{j}=\mathbf{B}_{\text P}\mathbf{h}^{j}+\mathbf{q}_{\text P}^{j}, \quad j=1,2,\cdots,N,
\end{equation}
where
\[
\mathbf{y}_{\text P}^{j}=[y_{\text P}^{j}(0),y_{\text P}^{j}(1),\cdots,y_{\text P}^{j}(KL-1)]^T, \quad \quad
\mathbf{h}^{j}=[(\mathbf{h}^{(j,1)})^T,\cdots,(\mathbf{h}^{(j,k)})^T,\cdots,(\mathbf{h}^{(j,K)})^T]^T,
\]
\[
\mathbf{h}^{(j,k)}=[h^{(j,k)}(0),h^{(j,k)}(1),\cdots,h^{(j,k)}(L-1)]^T, \quad \quad
\mathbf{q}_{\text P}^{j}=[q_{\text P}^{j}(0),q_{\text P}^{j}(1),\cdots,q_{\text P}^{j}(KL-1)]^{T},
\]
\[
\mathbf{B}_{\text P}=[\mathbf{B}_{{\text p}_1} \mathbf{B}_{{\text P}_2}\cdots\mathbf{B}_{{\text P}_{K}}], \quad \quad
\mathbf{B}_{{\text P}_{k}}=[\mathbf{0}_{L\times (k-1)L}\quad b\mathbf{I}_L \quad \mathbf{0}_{L\times (K-k)L }]^T.
\]
We use $b=\sqrt{KE_s\left(\sum_{l=0}^{L-1}\Omega _{l}^2\right)}$ to maintain the same average receive SNR in both pilot phase and data phase.

From the signal observed at the $j$th receive antenna from time $0$ to $KL-1$
during pilot phase, we obtain an initial estimate the channel vector ${\bf h}^j$
using the scaled identity nature of $\mathbf{B}_{\text P}$, as
\begin{eqnarray}
\mathbf{\widehat{h}}^{j} = \mathbf{y}_{\text P}^{j}/b, \quad j=1,2,\cdots,N.
\label{init_channel_est}
\end{eqnarray}
These initial channel estimates are used to detect the data vectors in the data 
phase.

\subsection{Equalization using initial channel estimates}
\label{sec5c}
In the data phase, let
$\mathbf{a}_{i}^{k} =[{a}_{i}^{k}(0),{a}_{i}^{k}(1),\cdots,{a}_{i}^{k}(I+L-2)]^{T}$ 
denote the data vector of size $(I+L-1)\times 1$, which includes $(L-1)$ cyclic
prefix symbols and $I$ information symbols transmitted from $k$th user during $i$th 
data block, $i=1,2,\cdots,Q$. The signal received at $j$th receive antenna at $n$th
channel use of $i$th data block is given by
\begin{equation}
\label{eq7}
y_{i}^{j}(n)=\sum_{k=1}^{K}\sum_{l=0}^{L-1}h^{(j,k)}(l)a^{k}_{i}(n-l)+q_{i}^{j}(n),
 \end{equation}
$j=1,2\cdots,N,\, n=0,1,\cdots,I+L-2$, where ${q}_{i}^{j}(n)$ is the
noise sample modeled as i.i.d. $\mathcal{CN}(0,\sigma^2)$.
Define the following vectors and matrices:
$\mathbf{y}_{i}^{j}\Define [y_{i}^{j}(L-1),y_{i}^{j}(L),\cdots,y_{i}^{j}(I+L-2)]^{T}$,
$\mathbf{q}_{i}^{j}\Define [q_{i}^{j}(L-1),q_{i}^{j}(L),\cdots,q_{i}^{j}(I+L-2)]^{T}$,
$\mathbf{x}_{i}^{k}\Define [{a}_{i}^{k}(L-1),{a}_{i}^{k}(L),\cdots,{a}_{i}^{k}(I+L-2)]^{T}$, and
$\mathbf{H}^{j,k}$ as a $(I+L)\times I$ circulant matrix with
$[h^{(j,k)}(0),h^{(j,k)}(1),\cdots,h^{(j,k)}(L-1),0,\cdots,0]^T$ as
the first column. With these definitions, (\ref{eq7}) can be written
in the form
\begin{eqnarray}
\label{eq8}
\mathbf{y}_{i}^{j}=\sum_{k=1}^{K}\mathbf{H}^{j,k}\mathbf{x}_{i}^{k}+\mathbf{q}_{i}^{j},\quad j=1,2,\cdots,N.
\end{eqnarray}
We can write (\ref{eq8}) as
\begin{equation}
\label{eq9}
\mathbf{y}_{i}=\mathbf{H}\mathbf{x}_{i}+\mathbf{q}_{i}, \quad i=1,2,\cdots,Q,
\end{equation}
where
{\small
$\mathbf{y}_{i}=[(\mathbf{y}_{i}^{1})^{T},(\mathbf{y}_{i}^{2})^{T},\cdots,(\mathbf{y}_{i}^{N})^{T}]^{T}$,
$\mathbf{x}_{i}=[(\mathbf{x}_{i}^{1})^{T},(\mathbf{x}_{i}^{2})^{T},\cdots,(\mathbf{x}_{i}^{K})^{T}]^{T}$,
$\mathbf{q}_{i}=[(\mathbf{q}_{i}^{1})^{T},(\mathbf{q}_{i}^{2})^{T},\cdots,(\mathbf{q}_{i}^{N})^{T}]^{T}$}, 
and
\[
\mathbf{H}=\left[
\begin{array}{ccc}
\mathbf{H}^{1,1} & \mathbf{H}^{1,2}\cdots & \mathbf{H}^{1,K} \\
\mathbf{H}^{2,1} & \mathbf{H}^{2,2}\cdots & \mathbf{H}^{2,K} \\
\vdots & \vdots &\vdots\\
\mathbf{H}^{N,1} & \mathbf{H}^{N,2}\cdots & \mathbf{H}^{N,K} \\
\end{array}
\right].
\]

\subsubsection{Equalization using R-MCMC-R method}
\label{sec5c1}
The R-MCMC-R algorithm proposed in Sec. \ref{sec3} is employed in the frequency 
domain using FFT based processing for equalization. The circulant matrix 
$\mathbf{{H}}^{j,k}$ can be decomposed as
\begin{equation}
\mathbf{H}^{j,k}=\mathbf{F}_{I}^{H}\mathbf{D}^{j,k}\mathbf{F}_{I},
\end{equation}
where $\mathbf{F}_{I}$ is $I\times I$ DFT matrix, and $\mathbf{D}^{j,k}$
is a diagonal matrix with its diagonal elements to be the DFT of the vector
$[h^{(j,k)}(0),h^{(j,k)}(1),\cdots,h^{(j,k)}(L-1),0,\cdots,0]^T$. 
Taking the DFT of $\mathbf{{y}}_{i}^{j}$ in (\ref{eq8}), we get
\begin{equation}
\label{eq12}
\mathbf{z}_{i}^{j}=\mathbf{F}_{I}\mathbf{{y}}_{i}^{j}=\sum_{k=1}^{K}\mathbf{D}^{j,k}\mathbf{{b}}_{i}^{k}+\mathbf{w}_{i}^{j},\quad j=1,2,\cdots,N,
\end{equation}
where
$\mathbf{z}_{i}^{j}=[z_i^j(0),z_i^j(1),\cdots,z_i^j(I-1)]^T$,\,
$\mathbf{{b}}_{i}^{k}\Define \mathbf{F}_{I}\mathbf{x}_{i}^{k}
= [b_i^k(0),b_i^k(1),\cdots,b_i^k(I-1)]^T$,
and
$\mathbf{w}_{i}^{j} \Define \mathbf{F}_{I}\mathbf{{q}}_{i}^{j}
= [w_i^j(0),w_i^j(1),\cdots,w_i^j(I-1)]^T$.
Writing (\ref{eq12}) in matrix form, we get
\begin{equation}
\label{eq13}
\mathbf{z}_{i}=\mathbf{D}\mathbf{{b}}_{i}+\mathbf{w}_{i}, \quad i=1,2,\cdots,Q,
\end{equation}
where 
{\small $\mathbf{z}_{i}=[(\mathbf{z}_{i}^{1})^{T},(\mathbf{z}_{i}^{2})^{T},\cdots,(\mathbf{z}_{i}^{N})^{T}]^{T}$,
 $\mathbf{{b}}_{i}=[(\mathbf{{b}}_{i}^{1})^{T},(\mathbf{{b}}_{i}^{2})^{T},\cdots,(\mathbf{{b}}_{i}^{K})^{T}]^{T}$,
$\mathbf{w}_{i}=[(\mathbf{w}_{i}^{1})^{T},(\mathbf{w}_{i}^{2})^{T},\cdots,(\mathbf{w}_{i}^{N})^{T}]^{T}$}, 
and
\[
\mathbf{D}=\left[
\begin{array}{ccc}
\mathbf{D}^{1,1} & \mathbf{D}^{1,2}\cdots & \mathbf{D}^{1,K} \\
\mathbf{D}^{2,1} & \mathbf{D}^{2,2}\cdots & \mathbf{D}^{2,K} \\
\vdots & \vdots &\vdots\\
\mathbf{D}^{N,1} & \mathbf{D}^{N,2}\cdots & \mathbf{D}^{N,K} \\
\end{array}
\right].
\]
Rearranging the terms, we can also write (\ref{eq13}) as
\begin{equation}
\label{eq14}
\mathbf{\bar{z}}_{i}=\mathbf{\bar{D}}\mathbf{\bar{b}}_{i}+\mathbf{\bar{w}}_{i}, \quad i=1,2,\cdots,Q,
\end{equation}
where
\[
\mathbf{\bar{z}}_{i}=\left[
\begin{array}{c}
\mathbf{\bar{z}}_{i}(0)\\
\mathbf{\bar{z}}_{i}(1) \\
\vdots\\
\mathbf{\bar{z}}_{i}(I-1)
\end{array}\right]
\mathrm{,} \,\, 
\mathbf{\bar{b}}_{i}=\left[
\begin{array}{c}
\mathbf{\bar{b}}_{i}(0)\\
\mathbf{\bar{b}}_{i}(1) \\
\vdots\\
\mathbf{\bar{b}}_{i}(I-1)
\end{array}\right], \,\,
\mathbf{\bar{D}}=
\left[
\begin{array}{ccc}
\mathbf{\bar{D}}(0)  & \cdots & 0 \\
\vdots &  \ddots & \vdots \\
0  & \cdots & \mathbf{\bar{D}}(I-1) \\
\end{array}
\right]
\mathrm{,} \,\, 
\mathbf{\bar{w}}_{i}=\left[
\begin{array}{c}
\mathbf{\bar{w}}_{i}(0)\\
\mathbf{\bar{w}}_{i}(1) \\
\vdots\\
\mathbf{\bar{w}}_{i}(I-1)
\end{array}\right],
\]
$\mathbf{\bar{z}}_{i}(m)=[{z}_{i}^{1}(m),{z}_{i}^{2}(m),\cdots,{z}_{i}^{N}(m)]^T$,
$\mathbf{\bar{b}}_{i}(m)=[{b}_{i}^{1}(m),{b}_{i}^{2}(m),\cdots,{b}_{i}^{N}(m)]^T$,
\\
$\mathbf{\bar{w}}_{i}(m)=[{w}_{i}^{1}(m),{w}_{i}^{2}(m),\cdots,{w}_{i}^{N}(m)]^T$,
and

\[
\mathbf{\bar{D}}(m)=\left[
\begin{array}{ccc}
\mathbf{D}^{1,1}(m) & \mathbf{D}^{1,2}(m)\cdots & \mathbf{D}^{1,K}(m) \\
\mathbf{D}^{2,1}(m) & \mathbf{D}^{2,2}(m)\cdots & \mathbf{D}^{2,K}(m) \\
\vdots & \vdots &\vdots\\
\mathbf{D}^{N,1}(m) & \mathbf{D}^{N,2}(m)\cdots & \mathbf{D}^{N,K}(m) \\
\end{array}
\right].
\]
$\mathbf{D}^{j,k}(m)$ is the $m$th diagonal element of the matrix
$\mathbf{D}^{j,k}$.
Also, $\mathbf{\bar{b}}_{i}=\mathbf{{\bar F}}\mathbf{\bar{x}}_{i}$, where
$\mathbf{{\bar F}}\Define \mathbf{F}_{I}\otimes \mathbf{I}_{K}$,
$\mathbf{\bar{x}}_{i}=[a_{i}^{1}(L-1)\cdots a_{i}^{K}(L-1),a_{i}^{1}(L)\cdots a_{i}^{K}(L),\cdots,a_{i}^{1}(I+L-2)\cdots a_{i}^{K}(I+L-2)]^T$.
Now, we have
\begin{eqnarray}
\label{eq15}
\mathbf{\bar{z}}_{i} & = & \mathbf{\bar{D}}\mathbf{{\bar F}}\mathbf{\bar{x}}_{i}+\mathbf{\bar{w}}_{i} \nonumber \\
& = &\mathbf{{\bar H}}\mathbf{\bar{x}}_i+\mathbf{\bar{w}}_i,\quad i=1,2,\cdots,Q,
\end{eqnarray}
where $\mathbf{{\bar H}}\Define \mathbf{\bar{D}}\mathbf{{\bar F}}$.

For each $i$ in (\ref{eq15}), we run R-MCMC-R detection algorithm and detect 
the information symbols in the $i$th block. In the first iteration of data 
detection, we use the channel estimates from (\ref{init_channel_est})  
to calculate $\mathbf{\widehat{{\bar  H}}}$, an estimate of $\mathbf{{\bar  H}}$.
Each coordinate of the vector 
$(\mathbf{\bar{z}}_{i}-\mathbf{\widehat{{\bar H}}}\mathbf{\bar{x}}_i)$ has 
zero mean and $2\sigma^2$ variance. Using this knowledge, the statistics of 
the ML cost of error-free vectors are recalculated in the R-MCMC-R algorithm.
R-MCMC-R detector outputs are denoted by $\hat{{\bf x}}_i^k$, $k=1,\cdots,K$, \, 
$i=1,\cdots,Q$. 
These output vectors are then used to improve the channel estimates through 
iterations between equalization and channel estimation. The channel estimation
in these iterations is based on MCMC approach presented next.

\subsection{MCMC based channel estimation in data phase}
\label{sec5c2}
Consider (\ref{eq12}), which can be rewritten as
\begin{equation}
\label{eq17}
\mathbf{z}_{i}^{j}=\sum_{k=1}^{K}\mathbf{{B}}_{i}^{k}\mathbf{d}^{j,k}+\mathbf{w}_{i}^{j},\quad j=1,2,\cdots,N,
\end{equation}
where $\mathbf{{B}}_{i}^{k}=diag(\mathbf{{b}}_{i}^{k})$, $\mathbf{d}^{j,k}$
is a vector consisting of the diagonal elements of matrix $\mathbf{D}^{j,k}$,
which is the $I$-point DFT of $\mathbf{h}^{(j,k)}$ (zero padded to length
$I$), i.e.,
$\mathbf{d}^{j,k}=\tilde{\mathbf{F}}_{I\times L}\mathbf{h}^{(j,k)}$,
where $\tilde{\mathbf{F}}_{I\times L}$ is the matrix with the first $L$
columns of $\mathbf{F}_I$. Now, (\ref{eq17}) can be written as
\begin{eqnarray}
\label{eq18}
\mathbf{z}_{i}^{j}=\sum_{k=1}^{K}\mathbf{{B}}_{i}^{k}\tilde{\mathbf{F}}_{I\times L}\mathbf{h}^{(j,k)}+\mathbf{w}_{i}^{j}.
\end{eqnarray}
Defining 
$\mathbf{{A}}_{i}^{k}\Define \mathbf{{B}}_{i}^{k}\tilde{\mathbf{F}}_{I\times L}$,
we can write (\ref{eq18}) as
\begin{equation}
\label{eq19}
\mathbf{z}_{i}^{j}=\mathbf{{A}}_{i}\mathbf{h}^{j}+\mathbf{w}_{i}^{j}, \quad i=1,\cdots,Q,
\end{equation}
where  $\mathbf{A}_i = [\mathbf{{A}}_{i}^{1} \, \mathbf{{A}}_{i}^{2} \, \cdots \, \mathbf{{A}}_{i}^{K}]$.
We can write (\ref{eq19}) as
\begin{equation}
\label{eq20}
\mathbf{z}^{j}=\mathbf{{A}}\mathbf{h}^{j}+\mathbf{w}^{j},
\end{equation}
where
\[
\mathbf{{z}}^{j}=\left[
\begin{array}{c}
\mathbf{{z}}_{1}^{j}\\
\mathbf{{z}}_{2}^{j} \\
\vdots\\
\mathbf{{z}}_{Q}^{j}
\end{array}\right], \,\,
\mathbf{{A}}=\left[
\begin{array}{c}
\mathbf{{A}}_{1}\\
\mathbf{{A}}_{2} \\
\vdots\\
\mathbf{{A}}_{Q}
\end{array}\right], \,\,
\mathbf{{w}}^{j}=\left[
\begin{array}{c}
\mathbf{{w}}_{1}^{j}\\
\mathbf{{w}}_{2}^{j} \\
\vdots\\
\mathbf{{w}}_{Q}^{j}
\end{array}\right].
\]
Using the signal received at antenna $j$ from blocks $1$ to $Q$ in a frame
(i.e., using $\mathbf{z}^{j}$) and the matrix $\mathbf{\hat{A}}$ which is
formed by replacing the information symbols $\{\mathbf{x}_i^k\}$ in
$\mathbf{A}$ by the detected information symbols $\{\hat{\mathbf{x}}_i^k\}$,
the channel coefficients \{$\mathbf{h}^{j}$\} are estimated using MCMC based 
estimation technique presented in Sec. \ref{sec4d1}.
This ends one iteration between channel estimation and detection.
The detected data matrix is fed back for channel estimation in the next 
iteration, whose output is then used to detect the data matrix again.
This iterative channel estimation and detection procedure is carried out 
for a certain number of iterations. 

\subsection{Performance Results }
\label{sec5e}
\begin{figure}
\center
\includegraphics[totalheight=9.0cm,width=12.0cm]{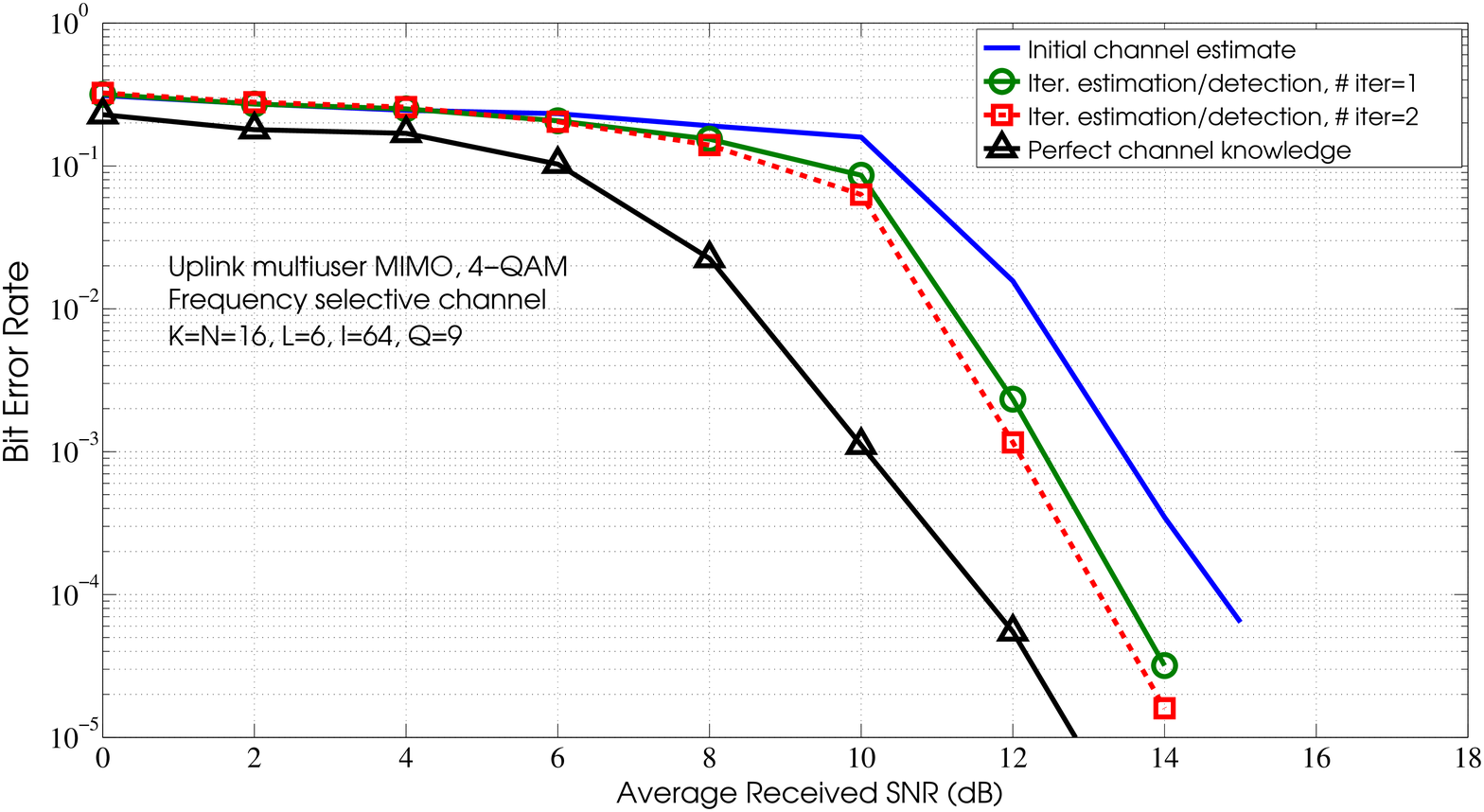}
\vspace{-6mm}
\caption{BER performance of iterative channel estimation/detection using MCMC 
based channel estimation and R-MCMC-R detection in uplink multiuser MIMO systems
on frequency selective fading with $K=N=16, L=6, I=64, Q=9$, 4-QAM. }
\label{fig16}
\vspace{-4.0mm}
\end{figure}

In Fig. \ref{fig16}, we plot the BER performance of the iterative channel 
estimation/detection scheme using proposed MCMC based channel estimation and 
R-MCMC-R detection algorithms in uplink multiuser MIMO system on frequency 
selective fading with $K=N=16, L=6, I=64, Q=9$ and 4-QAM. For the same 
settings, we also plot the BER performance of the R-MCMC-R algorithm with 
perfect channel knowledge. BER improves with increasing number of iterations
between channel estimation and detection. It can be observed that the proposed 
scheme with iterations is able to achieve performance which is close to the 
performance with perfect channel knowledge.

\section{Conclusions}
\label{sec6}
We proposed novel MCMC based detection and channel estimation algorithms that 
achieved near-optimal performance on the uplink in large-scale multiuser MIMO 
systems. The proposed R-MCMC-R detection algorithm was shown to alleviate the 
stalling problem and achieve near-ML performance in large systems with tens to 
hundreds of antennas and higher-order QAM. Key ideas that enabled such attractive
performance and complexity include $i)$ a randomized sampling strategy that gave the 
algorithm opportunities to quickly exit from stalled solutions and move to better 
solutions, and $ii)$ multiple random restarts that facilitated the algorithm to 
seek good solutions in different parts of the solution space. Multiple restarts 
alone (without randomized sampling) could not achieve near-ML performance at low 
complexity. Randomized sampling alone (without multiple restarts) could achieve 
near-ML performance  at low complexity in the case of 4-QAM. But for higher-order 
QAM (16-/64-QAM) randomized sampling alone was not adequate. Joint use of both 
randomized sampling as well as multiple restarts was found to be crucial to achieve 
near-ML performance for 16-/64-QAM. We also proposed an MCMC based channel
estimation algorithm which, in an iterative manner with the R-MCMC-R detection, 
achieved performance close to performance with perfect channel knowledge. We 
employed the proposed MCMC receiver architecture in the frequency domain for 
receiving CPSC signals on frequency selective fading between users and the BS.
While simulations were used to establish the attractiveness of the algorithm in 
performance and complexity, a theoretical analysis that could explain its good 
performance is important and challenging, which is a topic for future work. We 
have considered perfect synchronization and single-cell scenario in this paper. 
Other system level issues including uplink synchronization and multi-cell 
operation in large-scale MIMO systems can be considered as future work.

\newpage 

\newpage 

\begin{table*}[t]
\centering
\vspace{2cm}
\begin{tabular}{|c||c||c|c||c|c|}
\hline
 &  & \multicolumn{4}{|p{7.5cm}|}{\footnotesize Complexity in average number of real operations in $\times 10^{6}$ and SNR required to achieve $10^{-2}$ BER} \\ \cline{3-6}
Modulation  & Algorithm &  \multicolumn{2}{|c||}{$K=N=16$ } & \multicolumn{2}{|c|}{$K=N=32$ } \\
\cline{3-6}
& & Complexity & SNR & Complexity & SNR \\
\hline
\multirow{2}{*}{4-QAM} & R-MCMC-R (prop.) & 0.1424 & 9 dB & .848 & 8.8 dB  \\
\cline{2-6}
& R3TS \cite{r3ts} & 0.1877  & 9 dB & 0.6823 &8.8 dB \\
\cline{2-6}
& FSD in \cite{fsd} &  0.1351  & 10.1 dB & 4.9681  & 10.3 dB \\
\hline
\multirow{2}{*}{16-QAM}  & R-MCMC-R (prop.) &  1.7189  & 17 dB   & 15.158 & 16.7 dB \\
\cline{2-6}
& R3TS \cite{r3ts} &3.968 & 17 dB &  7.40464 &17 dB \\
\cline{2-6}
& FSD \cite{fsd} & 4.836432 & 17.6 dB   & 4599.5311 &  17.8 dB \\
\hline
\multirow{2}{*}{64-QAM}  & R-MCMC-R (prop.) &  11.181  & 24 dB & 166.284 &  24 dB \\
\cline{2-6}
& R3TS \cite{r3ts} &25.429504 & 24.2 dB & 77.08784 & 24.1 dB
\\
\cline{2-6}
& FSD in \cite{fsd} & 305.7204 & 24.3 dB & $\star$  &  $\star$  \\
\hline
\end{tabular}
\caption{Performance and complexity comparison of proposed R-MCMC-R detector
with other detectors in \cite{r3ts} and \cite{fsd} for for $K=N=16,32$ and 
4-/16-/64-QAM. $\star:$ Not simulated due to prohibitive complexity.
}
\label{tab1}
\end{table*}

\end{document}